\documentclass{article}
\usepackage{amssymb, amsmath, amsthm}
\usepackage{mathrsfs}
\usepackage{tabulary}
\usepackage{rotating}
\usepackage{enumerate}
\usepackage{setspace}
\usepackage{comment}
\usepackage{multirow}
\usepackage{paralist}
\usepackage{listings}
\usepackage[T1]{fontenc}
\usepackage[utf8]{inputenc}
\usepackage[english]{babel}
\usepackage{authblk}
\usepackage{etoolbox}
\usepackage{hyperref}
\usepackage{longtable}
\usepackage[usenames,dvipsnames]{color}

\newcommand\BibTeX{{\rmfamily B\kern-.05em \textsc{i\kern-.025em b}\kern-.08em
T\kern-.1667em\lower.7ex\hbox{E}\kern-.125emX}}

\usepackage{graphicx}
\graphicspath{{./figures/}}

\usepackage[style=nature]{biblatex}

\usepackage{caption}

\IfFileExists{upquote.sty}{\usepackage{upquote}}{}
\usepackage{hyperref}
\usepackage{subcaption}
\usepackage[hmargin=2.54cm,vmargin=2.54cm]{geometry}

\DeclareMathOperator{\expit}{expit}

\DeclareMathOperator{\EIC}{EIC}

\newcommand{\I}{\text{I}}

\usepackage{booktabs}
\usepackage{makecell} 
\usepackage{pgf,tikz}
 \usetikzlibrary{arrows,shapes.arrows,shapes.geometric,shapes.multipart,decorations.pathmorphing,positioning,shapes.swigs,calc,arrows.meta,fit,automata}

\usepackage{array}
\newcolumntype{L}[1]{>{\raggedright\let\newline\\\arraybackslash\hspace{0pt}}m{#1}}

\newcommand{\indep}{\perp \!\!\! \perp}

\usepackage{floatrow}
\DeclareFloatFont{tiny}{\tiny}
\usepackage{graphicx}
\begin{document}







\noindent Article Type: Original Research Article\\~\\

\noindent Title: Transporting results from a trial to an external target population when trial participation impacts adherence\\~\\

\noindent Authors: Rachael K. Ross, Iv\'an D\'iaz, Amy J. Pitts, Elizabeth A. Stuart, Kara E. Rudolph\\~\\

\noindent Corresponding author: Dr. Rachael Ross, Department of Epidemiology, Mailman School of Public Health, Columbia University, 722 W 168th St, New York, NY 10032; rachael.k.ross@columbia.edu\\~\\

\noindent Affiliations: Department of Epidemiology, Mailman School of Public Health, Columbia University, New York, NY (Ross, Rudolph); Department of Population Health, Grossman School of Medicine, New York University, New York, NY (D\'iaz), Department of Biostatistics, Mailman School of Public Health, Columbia University, New York, NY (Pitts); Department of Biostatistics, Bloomberg School of Public Health, Johns Hopkins University, Baltimore, MD (Stuart).\\~\\

\noindent Running header: Transporting when trial activities impact adherence\\~\\

\noindent Conflicts: None\\~\\

\noindent Funding: NIH NIDA R01DA056407\\~\\

\noindent Code is available at \url{https://github.com/rachael-k-ross/TransportSA}. Data used in the application are available at \url{https://datashare.nida.nih.gov/study/nida-ctn-0051} and \url{https://www.samhsa.gov/data/data-we-collect/teds-treatment-episode-data-set/datafiles}\\~\\

\noindent Word count: 4000/4000


\newpage
\section*{Abstract}
249/250 words\\~\\

\noindent Randomized clinical trials are considered the gold standard for informing treatment guidelines, but results may not generalize to real-world populations. Generalizability is hindered by distributional differences in baseline covariates and treatment-outcome mediators. Approaches to address differences in covariates are well established, but approaches to address differences in mediators are more limited. Here we consider the setting where trial activities that differ from usual care settings (e.g., monetary compensation, follow-up visits frequency) affect treatment adherence. When treatment and adherence data are unavailable for the real-world target population, we cannot identify the mean outcome under a specific treatment assignment (i.e., mean potential outcome) in the target. Therefore, we propose a sensitivity analysis in which a parameter for the relative difference in adherence to a specific treatment between the trial and the target, possibly conditional on covariates, must be specified. We discuss options for specification of the sensitivity analysis parameter based on external knowledge including setting a range to estimate bounds or specifying a probability distribution from which to repeatedly draw parameter values (i.e., use Monte Carlo sampling). We introduce two estimators for the mean counterfactual outcome in the target that incorporates this sensitivity parameter, a plug-in estimator and a one-step estimator that is double robust and supports the use of machine learning for estimating nuisance models. Finally, we apply the proposed approach to the motivating application where we transport the risk of relapse under two different medications for the treatment of opioid use disorder from a trial to a real-world population.
\\~\\

\noindent Keywords: transportability, adherence, target population, sensitivity analysis

\newpage
\doublespacing
\section*{Introduction}

Randomized clinical trials are considered the gold standard for producing evidence for treatment guidelines, however, the sample of patients included in trials is not often representative of the real-world target population who may be treated.\cite{rudolph_under-representation_2022} When the distribution of patient characteristics in the trial sample and in the real-world target are different and these patient characteristics affect the outcome, trial results (e.g., the mean potential outcome) are not naively (i.e., trivially \cite{pearl_transportability_2011}) transportable to the target. Causal inference approaches have been developed to formally transport treatment effects by accounting for (i.e., standardizing by) these differences \cite{cole_generalizing_2010,lesko_generalizing_2017,dahabreh_extending_2020,rudolph_robust_2017}. However, transportability of trial results is also compromised by differing distributions of mediators of the treatment assignment-outcome relationship like adherence.\cite{dahabreh_generalizing_2022,frangakis_calibration_2009,rudolph_robust_2017} For example, aspects of care in clinical trials may differ from usual-care settings, such as providing monetary compensation, nature and frequency of follow-up visits, and composition of the care team. Such differences are likely to impact adherence which in-turn impacts the outcome.\cite{wang_emulation_2023,franklin_emulating_2021} Although it has been recognized that transportability is compromised by differences in post-treatment mediators like adherence,\cite{pearl_transportability_2011,dahabreh_generalizing_2022,frangakis_calibration_2009,rudolph_robust_2017}, most applications have ignored this issue. Ignoring this issue is equivalent to transporting the effect of treatment assignment assuming trial activities do not affect the outcome or transporting the joint effect of treatment assignment \textit{and} the trial activities (i.e., as if these activities are included with treatment in the usual-care setting),\cite{dahabreh_generalizing_2022} neither of which may be realistic or answer the research question.

In this paper, we consider the setting where trial activities affect adherence to assigned treatment; and where data on treatment assignment and adherence are available in the trial, but not the target. We develop methods for a sensitivity analysis to examine sensitivity of results to the impact of trial activities on adherence by leveraging a user-specified parameter for the relative difference in adherence between the trial and target. We introduce estimators, discuss specification of the sensitivity analysis parameter, and apply the proposed approach in the motivating application in which we aimed to inform the choice between two medications, extended-release naltrexone (XR-NTX) and buprenorphine-naloxone (BUP-NX), for opioid use disorder (OUD) in usual-care settings. Therefore, we sought to transport the risk of relapse under assignment to each medication from a randomized trial to a real-world target.  

\section*{Notation}
Consider two distinct samples, a sample of individuals enrolled in a trial and a simple random sample of individuals from a real-world target population (i.e., non-nested trial design\cite{dahabreh_causal_2024}). Let $S$ be an indicator of trial participation (i.e., $S=1$ for individuals in the trial, $S=0$ for individuals in the target sample). Let $W$ be a vector of baseline covariates, $A$ a discrete time-fixed indicator of treatment assignment that was randomized in the trial, $Z$ a binary indicator of adherence to assigned treatment (e.g., if the adherence measure is initiation, as in our motivating application, then, for an individual assigned $a$, $Z=1$ if the individual initiated $a$, otherwise $Z=0$), and $Y$ the outcome measured at the end of follow-up. In the main text, we consider a single binary $Z$ at one time point. We consider adherence at multiple time points (or multiple binary measures of adherence at one time time point) in the Appendix. For all $n_1$ individuals in the trial, indexed by $i$, we observe the data vector $(W_i,S_i=1,A_i,Z_i,Y_i)$; for all $n_0$ individuals in the target, we observe the data vector $(W_i,S_i=0)$; $n=n_0+n_1$. 
Finally, for any random variable $X$, let $X^a$ denote the potential value of $X$ under treatment assignment $a$ (e.g., $Z^a$ is an indicator of whether an individual would be adherent to assignment if they had been assigned treatment $a$, potentially counter to fact).

\section*{Estimand and identification}
Our causal estimand is the mean potential outcome under $a$ in the target sample: $E[Y^a|S=0]$. (Under imperfect adherence, the mean potential outcome under adherence, $E[Y^{a,z=1}|S=0]$, is also of interest though we do not consider that estimand as it is discussed elsewhere.\cite{rudolph_robust_2017}) Figure \ref{fig:dags} includes the Directed Acyclic Graphs (DAGs) and the corresponding Single World Intervention Graphs (SWIGs) with an intervention on $A$ (as in our estimand) for two settings: i) trial participation only affects the outcome through treatment assignment and ii) trial participation also affects the outcome through adherence, depicted by an arrow from $S$ to $Z$ (from $S$ to $Z^a$ in the SWIG). Under setting (i), the probability of adherence to $a$ conditional on $W$ is equal in the trial and the target, $\Pr(Z=1|A=a,W,S=1)=\Pr(Z=1|A=a,W,S=0)$. 
In contrast, under setting (ii), this equality does not hold, $\Pr(Z=1|A=a,W,S=1)\neq\Pr(Z=1|A=a,W,S=0)$. 
We define the parameter $\delta_a(W)$ as the ratio of the potential adherence to $a$ in the target to potential adherence in the trial conditional on $W$, 
\begin{equation}
    \Pr(Z^a=1|W,S=0)=\Pr(Z^a=1|W,S=1)\delta_a(W).
    \label{delta}
\end{equation} 
Importantly we do not include an arrow from $S$ directly to $Y$. In this work we only consider potential indirect affects of trial activities on the outcome through an exposure-outcome mediator; we do not consider direct effects on the outcome. Note that in the transportability literature the direction of the arrow between $S$ and $W$ varies (e.g., see \cite{dahabreh_generalizing_2022,pearl_transportability_2011}). While we include $S\rightarrow W$, our work holds when $S\leftarrow W$. 

As previously shown, $E[Y^a|S=0]=E_W[E[Y|A=a,W,S=1]|S=0]$ under the following assumptions,\cite{dahabreh_extending_2020} which hold in setting (i): 
\begin{enumerate}
    \item[A1.] Mean conditional exchangeability over the samples: trial participation is independent of the mean potential outcome, conditional on covariates, $E(Y^a|W,S=1)-E(Y^a|W,S=0)$. 
    \item[A2.] Positivity of trial participation: there is a non-zero probability of trial participation across the distribution of $W$ in the target, $\Pr(S=1|W=w)>0$ when $\Pr(S=0|W=w)>0$. 
    \item[A3.] Conditional exchangeability over treatment assignment in the trial: in the trial, treatment assignment is independent of the potential outcome, conditional on covariates, $Y^a\indep A|W,S=1$. 
    \item[A4.] Positivity of treatment assignment in the trial: in the trial, there is a non-zero probability of being assigned to each treatment across the distribution of $W$ in the target, $\Pr(A=a|W=w,S=1)>0$ for all $a$ when $\Pr(S=0|W=w)>0$. 
    \item[A5.] Causal consistency in the trial: in the trial, an individual's observed value of post-treatment variables are the potential values under the individual's observed treatment assignment, i.e., $Z_i=Z_i^a$ and $Y_i=Y_i^a$ when $A=a$.
\end{enumerate}
Assumptions A3-A4 hold by design in the trial by randomization. 
We apply the law of total probability to illustrate how treatment adherence in the trial is implicitly included,
\begin{align}
    E[Y^a \mid S=0] &= E[E[Y|A=a,W,S=1]|S=0] \nonumber\\
       \begin{split}
    &= E\big[\big\{E[Y|A=a,W,Z=1,S=1]\Pr(Z=1|A=a,W,S=1)  \\
    &\quad\quad\quad\quad+ E[Y|A=a,W,Z=0,S=1]\left(1-\Pr(Z=1|A=a,W,S=1)\right) \big\}\mid S=0\big].
    \end{split} \label{standardeqn}
\end{align}
The interior portion of the estimand, $E[Y|A=a,W,S=1]$, is presented in (\ref{standardeqn}) as a weighted average: the expectation of the outcome under adherence and non-adherence weighted by the probabilities of adherence and non-adherence, respectively, \textit{in the trial}. That is, the estimand is a function of the adherence observed in the trial. Thus when trial participation affects adherence as in setting (ii), $E[Y^a|S=0]\neq E_W[E[Y|A=a,W,S=1]|S=0]$. Specifically, assumption A1 does not hold because of the open path $S\rightarrow Z^a \rightarrow Y^a$.

Thus we consider alternative versions of A1-A4 that hold in setting (ii), 
\begin{enumerate}
    \item[A1*.] Mean conditional exchangeability over the samples: participation in the trial is independent of the mean potential outcome, conditional on covariates \textit{and adherence}, $E(Y^a|W,Z^a,S=1)=E(Y^a|W,Z^a,S=0)$
    \item[A2*.] Positivity of trial participation: there is a non-zero probability of trial participation across the distribution of $W$ and $Z^a$ in the target, $\Pr(S=1|W=w,Z^a=z)>0$ for all $a$ when $\Pr(S=0|W=w,Z^a=z)>0$ 
    \item[A3*.] Conditional exchangeability over treatment assignment in the trial: in the trial, treatment assignment is independent of the potential outcome \textit{and potential adherence}, conditional on covariates, $(Y^a,Z^a) \indep A|W,S=1$
    \item[A4*.] Positivity of treatment assignment in the trial: in the trial, there is a non-zero probability of being assigned to each treatment across the distribution of $W$ and $Z^a$ in the target,
    $\Pr(A=a|W=w,Z^a=z,S=1)>0$ when $\Pr(S=0|W=w,Z^a=z)>0$ 
\end{enumerate}
Again, A3* and A4* hold by design in the trial. Under assumptions A1*-A4* plus A5, and using our definition of $\delta_a(W)$ (proof in Appendix),
\begin{align}
   \begin{split}
     E[Y^a|S=0]=E&\big[\big\{E\left( Y\mid A=a,Z=1,W,S=1\right)\Pr(Z=1|A=a,W,S=1)\delta_a(W) \\
   & + E\left( Y\mid A=a,Z=0,W,S=1\right)(1-\Pr(Z=1|A=a,W,S=1)\delta_a(W))\big\}\mid S=0\big].   \label{withdelta}
   \end{split}
\end{align}

Comparing equations (\ref{standardeqn}) and (\ref{withdelta}), the former only includes adherence from the trial while the latter captures the relative difference in adherence between the trial and the target. When $\delta_a(W)=1$, the trial and target conditional adherence are equal and equation (\ref{withdelta}) equals equation (\ref{standardeqn}). In practice, researchers commonly use (\ref{standardeqn}) implying or explicitly drawing a diagram like setting (i), thus making the assumption that $\delta_a(W)=1$. Our proposed alternative (\ref{withdelta}) allows one to examine the sensitivity of results to this assumption. $\delta_a(W)$ is unknown and cannot be identified with the data, so this parameter must be specified for estimation (discussed below). In the main text we consider only baseline covariates $W$; we consider settings with post-treatment assignment covariates, $L$ in the Appendix. 

When one is interested \textit{only} in differences in mean potential outcomes under different values of $a$, $E[Y^a-Y^{a'}|S=0]$
, then A1 and A1* can be replaced with weaker versions: conditional exchangeability \textit{in measure} over $S$.\cite{dahabreh_extending_2020} For example, for A1, this is $E(Y^a-Y^{a'}|W,S=1)=E(Y^a-Y^{a'}|W,S=0)$, which is satisfied when $W$ contains the variables that modify the effect difference (a subset of all the variables that effect the outcome) and differ in distribution across the trial and target. 

\section*{Estimation and inference}
We provide two estimators for (\ref{withdelta}), henceforth $\psi(a,\delta_a(W))$: a g-computation estimator, $\widehat\psi_{G}(a,\delta_a(W))$, and a one-step estimator based on the efficient influence curve (EIC), $\widehat\psi_{OS}(a,\delta_a(W))$. We also discuss estimators for the sampling variance for constructing confidence intervals. Throughout we treat $\delta_a(W)$ as a known parameter that does not have sampling error. We define additional notation for conciseness: 
\begin{itemize}
    \item The $W$-conditional outcome expectation when $A=a$ and $Z=z$ in the trial: $Q_{a,z}(W)=E[Y|A=a,Z=z,W,S=1]$
    \item The $W$-conditional probability of adherence when $A=a$ in the trial: $m_a(W)=\Pr(Z=1|A=a,W,S=1)$
    \item The $W$-conditional probability of treatment assignment $a$ in the trial: $g_a(W)=\Pr(A=a|W,S=1)$
    \item The $W$-conditional probability of selection into the trial: $h(W)=\Pr(S=1|W)$
\end{itemize}
Additionally $\I(\cdot)$ is the identity function that equals $1$ when true and $0$ otherwise.

\subsection*{Estimators}

\subsubsection*{G-computation estimator}
This estimator is a plug-in estimator based on (\ref{withdelta}),
$$\widehat\psi_{G}(a,\delta_a(W))=
\frac{1}{n_0}
\sum_{i=1}^n \I(S_i=0) \left\{\widehat Q_{a,1}(W_i) \widehat m_a(W_i) \delta_a(W) + \widehat Q_{a,0}(W_i)(1- \widehat m_a(W_i) \delta_a(W))\right\}.$$
This estimator includes two nuisance models, $\widehat Q_{a,z}(W)$ and $\widehat m_a(W)$, and consistency of $\widehat\psi_{G}(a,\delta_a)$ depends on correct specification of both models. The likelihood of correctly specifying these models is increased by using machine learning, however, valid inference (i.e., confidence intervals) is not theoretically supported when using machine learning with $\widehat\psi_{G}(a,\delta_a)$.\cite{renson_pulling_2025} This motivates our derivation of the one-step estimator that is based on the EIC and thus has asymptotic properties that support valid inference with machine learning.\cite{renson_pulling_2025}

\subsubsection*{One-step estimator}
The conjectured EIC of (\ref{withdelta}) is (derivation in Appendix)\cite{kennedy_semiparametric_2023} 
{\small{
\begin{align*}
\phi&=\frac{1}{\Pr(S=0)} \Bigg[\frac{\I(A=a)S}{g_a(W)}\frac{1-h(W)}{h(W)}\\
&\quad\quad\quad\bigg\{\bigg(Z\delta_a(W) + (1-Z)\frac{(1-m_a(W)\delta_a(W))}{1-m_a(W)}\bigg)\left(Y-Q_{a,Z}(W)\right)  + \delta_a(W)\left(Q_{a,1}(W)- Q_{a,0}(W)\right)\left(Z-m_a(W)\right) \bigg\}\\
&\quad\quad\quad+ (1-S)\left\{ Q_{a,1}(W)m_a(W)\delta_a(W) + Q_{a,0}(W)(1- m_a(W)\delta_a(W))- \psi(a,\delta_a(W)) \right\}\Bigg].
\end{align*}}}
\noindent The one-step estimator is derived by adding the sample average of the estimated EIC to $\widehat\psi_{G}(a,\delta_a)$\cite{hines_demystifying_2022},
{\small
\begin{align*}
\widehat\psi_{OS}&(a,\delta_a(W))\\
&=\frac{1}{n_0}\sum_{i=1}^n 
\Bigg[\frac{\I(A_i=a)S_i}{\widehat g_a(W_i)}\frac{1-\widehat h(W_i)}{\widehat h(W_i)}\\
&\quad\quad\quad\Bigg\{
\left(Z_i\delta_a(W) + (1-Z_i)\frac{1-\widehat 
 m_a(W_i)\delta_a(W)}{1-\widehat  m_a(W_i)} \right)
\left(Y_i - \widehat  Q_{a,Z_i}(W_i)\right)- \delta_a(W)\left(\widehat Q_{a,1}(W_i)- \widehat Q_{a,0}(W_i)\right)\left(Z_i-\widehat m_a(W_i)\right) \Bigg\}\\
&\quad\quad\quad+ (1-S_i)\left\{ \widehat Q_{a,1}(W_i)\widehat m_a(W_i)\delta_a(W) + \widehat Q_{a,0}(W_i)\left(1- \widehat m_a(W_i)\delta_a(W)\right)\right\}\Bigg].
\end{align*}}
This estimator includes four nuisance models, $\widehat Q_{a,z}(W)$, $\widehat m_a(W)$, $\widehat g_a(W)$ or $\widehat h(W)$. This estimator is double robust (i.e., only a subset of nuisance models must be correctly specified for consistency): 1) $\widehat Q_{a,z}(W)$ and $\widehat m_a(W)$, 2) $\widehat g_a(W)$, $\widehat h_a(W)$ and $\widehat m_a(W)$, or 3) $\widehat Q_{a,z}(W)$, $\widehat g_a(W)$, and $\widehat h_a(W)$. It also has asymptotic properties that allow use of machine learning.\cite{renson_pulling_2025} (proofs in Appendix) 

\subsection*{Inference}
With parametric (i.e., finite dimension) nuisance models, we can estimate the sampling variance with nonparametric bootstrap, the sandwich variance estimator (estimating equations are provided in Appendix), and the sample variance of the estimated EIC divided by $n$.\cite{daniel_double_2018} When using machine learning or model selection procedures, we can use the estimated EIC with cross-fitting\cite{hines_demystifying_2022} and a variation on the bootstrap.\cite{tran_robust_2018}  

\subsection*{Specifying $\delta_a(W)$}
$\delta_a$ is unknown and cannot be identified with the data in our setting, so we must specify it. It is often reasonable to assume $\Pr(Z^a=1|W,S=0)\leq\Pr(Z^a=1|W,S=1)$ (trial activities improve adherence),\cite{wang_emulation_2023,franklin_emulating_2021} meaning $0 \leq \delta_a(W) \leq 1$. With expert knowledge or external sources of information, we can posit realistic ranges or probability distributions for $\delta_a(W)$. To simplify, we can assume $\delta_a(W)$ does not depend on $W$ so that $\delta_a(W) \equiv \delta_a$. 

In practice, when we specify a range for $\delta_a$, we can estimate bounds.\cite{manski_nonparametric_1990} 
All values in the range are considered equally likely and the bounds create a region of potentially valid values of the estimand or contrast of estimands. If we believe that not all values in the range are equally likely, we can specify a probability distribution (or distributions) for $\delta_a$. Then we can select random draws from the distribution (Monte Carlo [MC] sampling) and estimate the estimand for each draw, resulting in many estimates that can be summarized. This is commonly done in probabilistic bias analyses.\cite{lash_applying_2009} 
Summarizing the collection of estimates can be done visually and quantitatively.\cite{lash_applying_2009} 

\section*{Application}
Extended-Release Naltrexone vs. Buprenorphine for Opioid Treatment (X:BOT) was an open-label randomized trial (NCT02032433) that aimed to compare the effectiveness of two OUD medications: XR-NTX and BUP-NX.\cite{lee_comparative_2018} Our objective is to transport the 24-week risk of relapse under assignment to each medication to a real-world target population represented in the Treatment Episode Data Set: Admission (TEDS-A), which captures data on admissions to substance use treatment facilities in the US
.\cite{noauthor_treatment_nodate} 

There are challenges to initiating XR-NTX (an opioid antagonist) and BUP-NX (a partial opioid agonist). Initiation can cause withdrawal symptoms (i.e., precipitated withdrawal) by out-competing opioids for opioid-receptors. To avoid this adverse effect, patients should abstain from opioids long enough (potentially a few hours) to be experiencing some withdrawal symptoms before initiating BUP-NX.\cite{noauthor_tip_2021,lee_comparative_2018} For XR-NTX, patients should be abstinent after withdrawal for $\geq$3 days and up to 10 days before initiation.\cite{noauthor_tip_2021,lee_comparative_2018,shulman_rapid_2024} Thus, XR-NTX initiation is more difficult than BUP-NX initiation. In X:BOT, 72\% of patients assigned to XR-NTX (204/283) and 94\% of patients assigned to BUP-NX (270/287) successfully initiated assigned medication during follow-up.\cite{lee_comparative_2018} These percentages are much higher than typically seen. For example, in another trial just 47\% of patients initiated XR-NTX and 73\% initiated BUP-NX.\cite{korthuis_hiv_2022} Given the challenges with initiation and that trial activities (such as monetary reimbursement for time, and longer and more frequent contact with providers and other members of the treatment team during the period of initiation) likely improved patients' chances of initiation of both treatments in X:BOT, we believe adherence to these medications in the real-world is lower than in the trial. 

\subsection*{Methods}

\subsubsection*{Trial and target samples}
Between 2014-2016, X:BOT enrolled 570 participants during a voluntary inpatient detoxification admission at eight community-based inpatient substance use disorder treatment sites in the US.\cite{lee_comparative_2018} Participants were $\ge$18 years old, diagnosed with OUD (by Diagnostic and Statistical Manual of Mental Disorders-5), recently used non-prescribed opioids, and were not pregnant. The primary outcome was relapse by 24 weeks defined as regular use of non-study opioids after day 20 post-randomization, defined as 4 consecutive opioid use weeks (i.e., $\ge$ 1 day of use with Timeline Followback method, positive urine toxicology, or not providing a urine sample) or 7 consecutive days of self-reported opioid use. We included all enrolled participants and the binary indicator of adherence ($Z$) was defined as initiation of the assigned medication by day 21. 

We used the TEDS-A from 2014-2016 as a representative cross-section of the target population. These data are collected by state administrative systems in all 50 states, DC, and Puerto Rico. The unit of the data are admissions, not patients; we treated admissions as independent. We included admissions to inpatient detoxification facilities for individuals aged 18-64 who reported heroin, non-prescription methadone, or other/synthetic opioid as their primary, secondary, or tertiary substance used. We excluded admissions for individuals whose reason for treatment was alcohol or cannabis dependence/abuse, or a mental health condition; who were pregnant; or who were referred for treatment by the court or criminal justice system. We also excluded admissions with missing covariate data (see next section). Data were not available from South Carolina in 2014 and 2015 or from Oregon in 2015 and 2016. Appendix Table 2 shows the states included in the final sample by year. 

\subsubsection*{Covariates, $W$}
We considered a set of covariates available in both TEDS-A and X:BOT. We included sociodemographics: age, sex, race, ethnicity, education, employment status, and housing status; characteristics of opioid use: any intravenous use and age at first use; and reported use of other substances: cannabis, cocaine or crack, amphetamines, and sedatives. Choice of these variables followed prior research examining effect modification in X:BOT.\cite{nunes_sublingual_2021}. Appendix Table 3 includes details of variable coding. Age was categorized into age groups (18-20, 21-24, etc.) in TEDS-A and as continuous in X:BOT. We imputed continuous age in TEDS-A as the mean age of the age group 
Regarding race, we believe patients self-classified race (i.e., self-identification through a closed-ended question)\cite{martinez_conceptualization_2023}, which we operationalized into 3 groups: Black, white, and other (including multiple races). Sample sizes in X:BOT did not support more granular coding of this "other" category. We conceptualize race as a social construct.\cite{jones_invited_2001} 

Appendix Table 1 shows missing covariate data in TEDS-A. After applying the inclusion and exclusion criteria there were 446,210 admissions in the cohort. There were six covariates with missingness: age at first opioid use (0.7\% missing), race (1.5\%), ethnicity (1.5\%), employment status (2.5\%), housing status (3.2\%), and education (3.5\%). In total, 8.3\% of admissions had any missingness and were excluded, resulting in 409,293 admissions.

\subsubsection*{Analysis}
We estimated the risk of relapse by 24 weeks for each medication, i.e., for $a\in \{0,1\}$ where $a=0$ for BUP-NX and $a=1$ for XR-NTX, and the risk difference (RD) using BUP-NX as the referent. First, we estimated the risks and risk difference \textit{in the X:BOT trial} (i.e., $E[Y^a|S=1]$)  standardized by the covariates, i.e., an adjusted version of the primary trial analysis.\cite{lee_comparative_2018} Second, we estimated the risks and risk difference \textit{in the TEDS-A cohort} assuming that trial activities did not affect adherence (i.e., $E[Y^a|S=0]$ as in setting (i), using assumptions A1-A5). For these two analyses we used one-step estimators previously described (Appendix).\cite{dahabreh_extending_2020,daniel_double_2018} 

Next, we estimated $E[Y^a|S=0]$ relying on assumptions A1*-A4* and A5, using the one-step estimator we derived above. We repeated the analysis under two approaches for specifying $\delta_a(W)$. For both, $\delta_a$ was not dependent on $W$, $\delta_a \equiv \delta(a)$, i.e.,  adherence ratios (one for $a=1$ and one for $a=0$) were considered constant across levels of the covariates. Note that a constant ratio does \textit{not} force adherence to be constant across levels of the covariates; just that the ratio of adherence between the trial and population was the same across values of the covariates. First, we considered static values 0.5 and 0.75 for both $\delta_1$ (for XR-NTX) and $\delta_0$ (for BUP-NX). Second, we specified distinct trapezoidal probability distributions for $\delta_1$ and $\delta_0$ (Appendix Figure 1). The distributions had the same minimum and maximum: $\{0.5,1\}$, but the modes differed: 0.6 and 0.75 for XR-NTX ($\delta_1$) and 0.75 and 0.9 for BUP-NX  ($\delta_0$). This difference reflects our belief, based on external knowledge, that adherence to XR-NTX is more impacted by trial activities than adherence to BUP-NX. We used MC sampling to estimate the risks and the RDs for 10000 repeated draws independent from each distribution. 

Outcome and adherence models were stratified by medication. All models were estimated using SuperLearner\cite{van_der_laan_super_2007} with a library including intercept-only model, main-effects logistic regression, gradient boosting machines, and multivariate adaptive regression splines, using the mlr3superlearner package.\cite{williams_mlr3superlearner_2024} The number of cross-validation folds was dynamically chosen based on the effective sample size for each model.\cite{phillips_practical_2023} We used cross-fitting with 30 folds. For inference, we estimated the sample variance of the estimated EIC divided by $n$ and constructed Wald-type 95\% confidence intervals.

\subsection*{Results}

\subsubsection*{Samples}
Table \ref{table:1} presents characteristics of the TEDS-A cohort and the X:BOT sample. There were 409,293 admissions in TEDS-A and 570 individuals in X:BOT. The TEDS-A cohort had a median age of 32 years, was 70\% male, 74\% white, and 16\% Hispanic/Latine. The X:BOT sample was similar on most sociodemographic characteristics, but individuals in X:BOT had more education (44\% with greater than high school versus 25\% in TEDS-A) and a smaller portion were unemployed (63\% versus 88\% in TEDS-A). 
The TEDS-A cohort and the X:BOT sample had similar characteristics of opioid use, but there were differences in the use of other substances with the use of other substances being notably higher in the X:BOT sample than in the TEDS-A cohort.


\subsubsection*{Adherence}
In X:BOT, 66.4\% (188/283) of patients assigned to XR-NTX and 94.1\% (270/287) of patients patients assigned BUP-NX initiated medication with 21 days. We used the estimated adherence model for each medication in the trial to examine the distribution of the predicted probabilities of adherence in the TEDS-A cohort (Figure \ref{fig:disadh}, Table \ref{table:adh}). Assuming that conditional adherence is the same in TEDS-A as in X:BOT for both medications (i.e., $\delta_0=\delta_1=1$, as in setting (i)), the percentage of patients predicted to adhere to XR-NTX and BUP-NX in TEDS-A was 64.3\% and 92.0\%, respectively. The individual-level variation in the predicted probability of adherence was larger for XR-NTX than BUP-NX (e.g., interquartile range: XR-NTX 0.587-0.707; BUP-NX 0.906-0.931). Assuming conditional adherence in TEDS-A is 50\% of that in X:BOT for both medications (i.e.,  $\delta_0=\delta_1=0.5$), the percentage of patients predicted to adhere to XR-NTX and BUP-NX in TEDS-A was 32.2\% and 46.0\%, respectively.


In Figure \ref{fig:draws} we plot the 10000 draws of $\delta_0$ and $\delta_1$ from the trapezoidal distributions and the resulting average predicted adherence for each medication in TEDS-A. Across all draws, the percentage of patients predicted to adhere ranged from 32.2\% to 63.9\% for XR-NTX (mean 46.3\%) and from 46.3\% to 91.7\% for BUP-NX (mean 72.0\%). For 3,427 (34.3\%) draws, the relative adherence parameter drawn for XR-NTX ($\delta_1$) was greater than the parameter drawn for BUP-NX ($\delta_0$) (gray dots in Figure \ref{fig:draws}). From external knowledge, we believe $\delta_1\leq\delta_0$, therefore we summarized results for all draws and the subset where $\delta_1\leq\delta_0$. 

\subsubsection*{Relapse}
In X:BOT, 61.1\% of individuals (348/570) relapsed by week 24. This percentage was 65.3\% among individuals randomized to XR-NTX and 56.8\% among individuals randomized to BUP-NX. Adjusting for baseline covariates, we estimated that the risk of relapse was 9.7 percentage points higher (95\% CI 1.2, 18.2) for XR-NTX than BUP-NX (\ref{table:results}).

First, we transported to the TEDS-A cohort assuming that adherence in TEDS-A would be the same as in X:BOT (assuming setting (i), i.e. $\delta_0=\delta_1=1$). We estimated the risk (as a percentage) of relapse was 79.8\% (95\% CI 60.9, 98.8) for XR-NTX and 48.8\% (95\% CI 24.9, 72.8) for BUP-NX, for a RD of 31.0 percentage points (95\% CI 0.4, 61.5) (Table \ref{table:results}). Next, we considered scenarios where adherence is lower in TEDS-A than in X:BOT, but $\delta_0=\delta_1$. As the $\delta$ values decreased, the risk of relapse increased, and the RD got larger. For example, at the lowest value considered, $\delta=0.5$, the risk difference was 34.1 percentage points (95\% CI 8.2, 60.1).

Finally, we repeated the analysis using the 10000 drawn values for $\delta_0$ and $\delta_1$ (bottom of Table \ref{table:results} and Figure \ref{fig:drawsresults}). Using all draws, the median risk of relapse was 85.7\% (simulation interval 81.0, 89.3) for XR-NTX and 51.4\% (48.8, 54.9) for BUP-NX; the median RD was 34.1 percentage points (28.4, 39.0). Restricting to draws where $\delta_1 \leq \delta_0$, the median RD was 35.4 percentage points (32.7, 39.4). 
 
\section*{Discussion}
We proposed a sensitivity analysis for studies that transport results from a trial to an external target population when trial activities may impact treatment adherence. Our approach leverages the adherence data that is only available in the trial and relies on specifying the relative difference in covariate-conditional adherence between the trial and the target. We could have alternatively defined a parameter for the additive difference in adherence, but using a relative measure ensures the adherence probabilities are bounded by 0 and 1. 

Sensitivity analyses for transportability analyses have been previously developed. Just like ours, these approaches assessed sensitivity of results to violations of the exchangeability over samples assumption (assumption A1).\cite{dahabreh_sensitivity_2023,huang_sensitivity_2022,nguyen_sensitivity_2017,nguyen_sensitivity_2018,zeng_efficient_2023} Our work is distinct, because we focused specifically on a violation that results when trial participation impacts adherence. In this case, we should not expect adherence in the target to be the same as in the trial. The previously proposed sensitivity analyses have considered bias resulting from violations of this exchangeability assumption generally (from one or multiple unspecified covariates)\cite{dahabreh_sensitivity_2023,zeng_efficient_2023} or due to a single unspecified covariate\cite{huang_sensitivity_2022,nguyen_sensitivity_2018}.

We used our proposed approach in an application estimating the transported risk of relapse under XR-NTX versus BUP-NX treatment for OUD. We found that accounting for the potential impact of trial activities on adherence increased the RD by between 3 and 6 percentage points, using various specifications of the sensitivity parameter. Our sensitivity analysis results were more precise than the results that assumed there was no impact of trial activities. These results indicate that the beneficial effect of BUP-NX compared to XR-NTX in preventing relapse is likely even greater in our real-world target population given our beliefs that adherence to both medications is worse in the target than the trial. 

Our work has limitations. We considered a single binary indicator for adherence. In the appendix we considered two time points (which could also be conceived as two binary measures of adherence at the same time point), but further work is needed to generalize the approach for longitudinal settings, settings where multiple (possibly high dimensional) aspects of adherence are impacted by trial activities. In our application we assumed the causal model drawn in setting (ii) where $Z$ was an indicator for initiation of the assigned medication. If participants with $Z=0$ were able to initiate an unassigned medication \textit{and} we believe trial activities impacted the distribution of other medication use, then we would need to include additional $Z$ nodes (i.e., multiple treatment-outcome mediators affected by $S$) or use a multi-valued adherence measure (e.g., $Z$ becomes an indicator for which medication was initiated). A multi-valued adherence measure would require specification of $\delta$ parameters for these additional values (for each treatment, for a $k$-valued adherence measure, we need $k-1$ parameters). Finally, in our application using MC sampling, we took draws from two independent probability distributions for $\delta_1$ and $\delta_0$ even though we had beliefs about the relationship of these parameters ($\delta_1\leq\delta_0$). We alternatively could have considered drawing from correlated distributions or drawing from a distribution for $\delta_1$ and a distribution for the ratio or difference between the parameters (e.g., $\delta_1/\delta_0)$. 

\singlespacing
\newpage
\printbibliography

\newpage
\section*{Tables}

\noindent \textbf{Table 1.} Characteristics of TEDS-A cohort and X:BOT sample
\begin{longtable}{lcc}
\label{table:1}
 & \multicolumn{2}{c}{\textbf{Sample, N(\%)}} \\ 
\cmidrule(lr){2-3}
& \textbf{TEDS-A} & \textbf{X:BOT} \\ 
\textbf{Characteristic}  & N = 409,293 &  N = 570\\
\midrule
\multicolumn{3}{l}{Sociodemographics} \\ 
\midrule
Age, median (IQR) & 32 (27, 42) & 31 (26, 39) \\ 
   18-20 & 12,119 (3.0\%) & 9 (1.6\%) \\ 
   21-24 & 53,520 (13\%) & 76 (13\%) \\ 
   25-29 & 98,324 (24\%) & 148 (26\%) \\ 
   30-34 & 77,862 (19\%) & 116 (20\%) \\ 
   35-39 & 50,196 (12\%) & 84 (15\%) \\ 
   40-44 & 35,595 (8.7\%) & 47 (8.2\%) \\ 
   45-49 & 33,415 (8.2\%) & 40 (7.0\%) \\ 
   50-54 & 26,313 (6.4\%) & 29 (5.1\%) \\ 
   55-64 & 21,949 (5.4\%) & 19 (3.3\%) \\ 
   65+ & 0 (0\%) & 2 (0.4\%) \\ 
Male & 285,632 (70\%) & 401 (70\%) \\ 
Race &  &  \\ 
   Black & 44,459 (11\%) & 57 (10\%) \\ 
   White & 301,741 (74\%) & 421 (74\%) \\ 
   Other & 63,093 (15\%) & 92 (16\%) \\ 
Hispanic/Latine & 63,614 (16\%) & 99 (17\%) \\ 
Education &  &  \\ 
   Less than high school & 107,654 (26\%) & 132 (23\%) \\ 
   High school or GED & 199,616 (49\%) & 190 (33\%) \\ 
   Greater than high school & 102,023 (25\%) & 248 (44\%) \\ 
Unemployed & 361,941 (88\%) & 360 (63\%) \\ 
Homeless & 102,616 (25\%) & 143 (25\%) \\ 
\midrule
\multicolumn{3}{l}{Opioid use} \\ 
\midrule
Intravenous use & 263,504 (64\%) & 385 (68\%) \\ 
Age at first use &  &  \\ 
      <15 & 25,477 (6.2\%) & 54 (9.5\%) \\ 
      15-20 & 162,998 (40\%) & 260 (46\%) \\ 
      20-29 & 143,977 (35\%) & 190 (33\%) \\ 
      30+ & 76,841 (19\%) & 66 (12\%) \\ 
\midrule
\multicolumn{3}{l}{Other substance use} \\ 
\midrule
Cannabis & 53,172 (13\%) & 305 (54\%) \\ 
Cocaine/crack & 88,156 (22\%) & 240 (42\%) \\ 
Amphetamines & 25,038 (6.1\%) & 128 (22\%) \\ 
Sedatives & 73,944 (18\%) & 203 (36\%) \\ 
\bottomrule
\end{longtable}
\noindent Abbreviations: TEDS-A, Treatment Episode Data Set: Admission; X:BOT, Extended-Release Naltrexone vs. Buprenorphine for Opioid Treatment trial; IQR, interquartile range

\newpage
\noindent \textbf{Table 2.} Observed adherence in X:BOT and predicted adherence in TEDS-A under different values of $\delta$
\begin{longtable}{lcc}
\label{table:adh}
& \textbf{XR-NTX} & \textbf{BUP-NX} \\ 
\midrule
\multicolumn{3}{l}{Observed in X:BOT} \\ 
\midrule
Total, N & 283 & 287 \\ 
Adherent ($Z=1$) & &\\ 
  N & 188 & 270 \\ 
  Probability & 0.664 & 0.941 \\ 
\midrule
\multicolumn{3}{l}{Predicted probability of adherence in TEDS-A\textsuperscript{a}}  \\
\midrule
Assuming no trial effect, i.e., $\delta_0=\delta_1=1$ & &\\
  Mean & 0.643 & 0.920\\
  Median (IQR)  & 0.632 (0.587, 0.707) & 0.911 (0.906, 0.931)\\
Assuming $\delta_0=\delta_1=0.75$ & &\\
  Mean & 0.482 & 0.690\\
  Median (IQR) & 0.474 (0.440, 0.530) & 0.684 (0.680, 0.698)\\
Assuming $\delta_0=\delta_1=0.5$ & &\\
  Mean & 0.322 & 0.460\\
  Median (IQR) & 0.316 (0.293, 0.354) & 0.456 (0.453, 0.465)\\
\bottomrule
\end{longtable}
\noindent Abbreviations: XR-NTX, extended-release naltrexone; BUP-NX, buprenorphine-naloxone; TEDS-A, Treatment Episode Data Set: Admission; X:BOT, Extended-Release Naltrexone vs. Buprenorphine for Opioid Treatment trial; IQR, interquartile range\\
\noindent \textsuperscript{a}Predicted using the nuisance model for adherence fit in the X:BOT sample

\newpage
\noindent \textbf{Table 3.} Estimated risks and risk differences in X:BOT trial and transported to TEDS-A cohort
\begin{longtable}{lccc}
\label{table:results}
 & \multicolumn{2}{c}{Risk as \% (95\% CI)} &   \\ 
 \cmidrule(lr){2-3}
 & XR-NTX & BUP-NX & Difference (95\% CI) \\ 
\midrule
In X:BOT trial\textsuperscript{a} & 66.2 (60.3, 72.0) & 56.4 (50.3, 62.5) & 9.7 (1.2, 18.2) \\ 
\midrule
Transported to TEDS-A cohort\textsuperscript{b} &  &  \\
\midrule
  Assuming no trial effect, i.e., $\delta_0=\delta_1=1$ & 79.8 (60.9, 98.8) & 48.8 (24.9, 72.8) & 31.0 (0.4, 61.5) \\
  Assuming $\delta_0=\delta_1=0.75$ & 84.9 (67.1, 1.0) & 52.1 (32.3, 71.9) & 32.8 (6.2, 59.5) \\
  Assuming $\delta_0=\delta_1=0.5$ & 90.1 (72.1, 1.0) & 56.9 (37.3, 74.7) & 34.1 (8.2, 60.1) \\
\\
\multicolumn{3}{l}{  MC sampling of $\delta$, Median (simulation interval\textsuperscript{c})} & \\
    All draws, n=10000 &&&\\
        Without random error & 85.7 (81.0, 89.3) & 51.4 (48.8, 54.9) & 34.1 (28.4, 39.0) \\
        Including random error\textsuperscript{d} & 85.4 (67.0, 1.0) & 51.8 (30.9, 71.9) & 33.8 (5.9, 61.9) \\
    Draws with $\delta_1 \leq \delta_0$, n=6573 &&&\\
        Without random error &  86.7 (82.5, 89.5) & 50.7 (48.6, 53.9) & 35.4 (32.7, 39.4) \\
        Including random error\textsuperscript{d} &  86.3 (68.2, 1.0) & 51.0 (29.8, 71.7) & 35.6 (7.5, 63.8) \\
\bottomrule
\end{longtable}
\noindent Abbreviations: XR-NTX, extended-release naltrexone; BUP-NX, buprenorphine-naloxone; TEDS-A, Treatment Episode Data Set: Admission; X:BOT, Extended-Release Naltrexone vs. Buprenorphine for Opioid Treatment trial; CI, confidence interval; MC, Monte Carlo\\
\noindent \textsuperscript{a}Adjusted for baseline covariates using a one-step estimator\\
\noindent \textsuperscript{b}Estimated using the proposed approach, except when we assumed no trial effect. For that scenario, we used a previously published one-step estimator\\
\noindent \textsuperscript{c}Simulation interval is the  2.5th and 97.5th percentiles of the point estimates\\
\noindent \textsuperscript{d}To incorporate random error to each estimate, we subtracted a random draw from a normal distribution with mean 0 and standard deviation set to the estimated standard error.\\

\newpage
\section*{Figures}
\noindent \textbf{Figure 1.} Directed acyclic graphs (DAG) and Single-World Intervention Graphs (SWIG)\\
\begin{figure}[h!]
\begin{flushleft}
 i)
\end{flushleft}  
\begin{center}
DAG
\begin{tikzpicture}
\tikzset{line width=1pt, outer sep=0pt, ell/.style={draw,fill=white, inner sep=2pt,line width=1pt}, swig vsplit={gap=5pt,inner line width right=0.5pt}};
\node[name=w, ell, shape=ellipse]{$W$};
\node[name=s, ell,above=5mm of w, shape=ellipse]{$S$};
\node[name=a, ell,right=5mm of w, shape=ellipse]{$A$};
\node[name=z, ell,right=5mm of a, shape=ellipse]{$Z$};
\node[name=y, ell,right=5mm of z, shape=ellipse]{$Y$};
\draw[->,line width=1pt,>=stealth]
(w) edge (s)
(w) edge (a)
(s) edge (a)
(a) edge (z)
(a) edge[out=45,in=145] (y)
(z) edge (y)
(w) edge[out=-60,in=-145] (y)
(w) edge[out=-45,in=-145] (z)
;
\end{tikzpicture}
SWIG              
\begin{tikzpicture}
\tikzset{line width=1pt, outer sep=0pt, ell/.style={draw,fill=white, inner sep=2pt,line width=1pt}, swig vsplit={gap=5pt,inner line width right=0.5pt}};
\node[name=w, ell, shape=ellipse]{$W$};
\node[name=s, ell,above=5mm of w, shape=ellipse]{$S$};
\node[name=a, ell,right=5mm of w, shape=swig vsplit]{
    \nodepart{left}{$A$}
    \nodepart{right}{$a$}};
\node[name=z, ell,right=5mm of a, shape=ellipse]{$Z^a$};
\node[name=y, ell,right=5mm of z, shape=ellipse]{$Y^{a}$};
\draw[->,line width=1pt,>=stealth]
(w) edge (s)
(w) edge (a)
(s) edge (a)
(a) edge (z)
(a) edge[out=45,in=145] (y)
(z) edge (y)
(w) edge[out=-60,in=-145] (y)
(w) edge[out=-45,in=-145] (z)
;
\end{tikzpicture}
\end{center}
\begin{flushleft}
ii)
\end{flushleft}  
\begin{center}
DAG
\begin{tikzpicture}
\tikzset{line width=1pt, outer sep=0pt, ell/.style={draw,fill=white, inner sep=2pt,line width=1pt}, swig vsplit={gap=5pt,inner line width right=0.5pt}};
\node[name=w, ell, shape=ellipse]{$W$};
\node[name=s, ell,above=5mm of w, shape=ellipse]{$S$};
\node[name=a, ell,right=5mm of w, shape=ellipse]{$A$};
\node[name=z, ell,right=5mm of a, shape=ellipse]{$Z$};
\node[name=y, ell,right=5mm of z, shape=ellipse]{$Y$};
\draw[->,line width=1pt,>=stealth]
(w) edge (s)
(w) edge (a)
(s) edge (a)
(s) edge (z)
(a) edge (z)
(a) edge[out=45,in=145] (y)
(z) edge (y)
(w) edge[out=-60,in=-145] (y)
(w) edge[out=-45,in=-145] (z)
;
\end{tikzpicture}
SWIG              
\begin{tikzpicture}
\tikzset{line width=1pt, outer sep=0pt, ell/.style={draw,fill=white, inner sep=2pt,line width=1pt}, swig vsplit={gap=5pt,inner line width right=0.5pt}};
\node[name=w, ell, shape=ellipse]{$W$};
\node[name=s, ell,above=5mm of w, shape=ellipse]{$S$};
\node[name=a, ell,right=5mm of w, shape=swig vsplit]{
    \nodepart{left}{$A$}
    \nodepart{right}{$a$}};
\node[name=z, ell,right=5mm of a, shape=ellipse]{$Z^a$};
\node[name=y, ell,right=5mm of z, shape=ellipse]{$Y^{a}$};
\draw[->,line width=1pt,>=stealth]
(w) edge (s)
(s) edge (z)
(w) edge (a)
(s) edge (a)
(a) edge (z)
(a) edge[out=45,in=145] (y)
(z) edge (y)
(w) edge[out=-60,in=-145] (y)
(w) edge[out=-45,in=-145] (z)
;
\end{tikzpicture}

\end{center}
\captionlistentry{}
\label{fig:dags}
\end{figure}

\newpage
\noindent \textbf{Figure 2.} Distribution of predicted probability of adherence to XR-NTX and BUP-NX in the TEDS-A cohort under different $\delta$ values\\
\begin{figure}[h!]
\includegraphics[width=\textwidth]{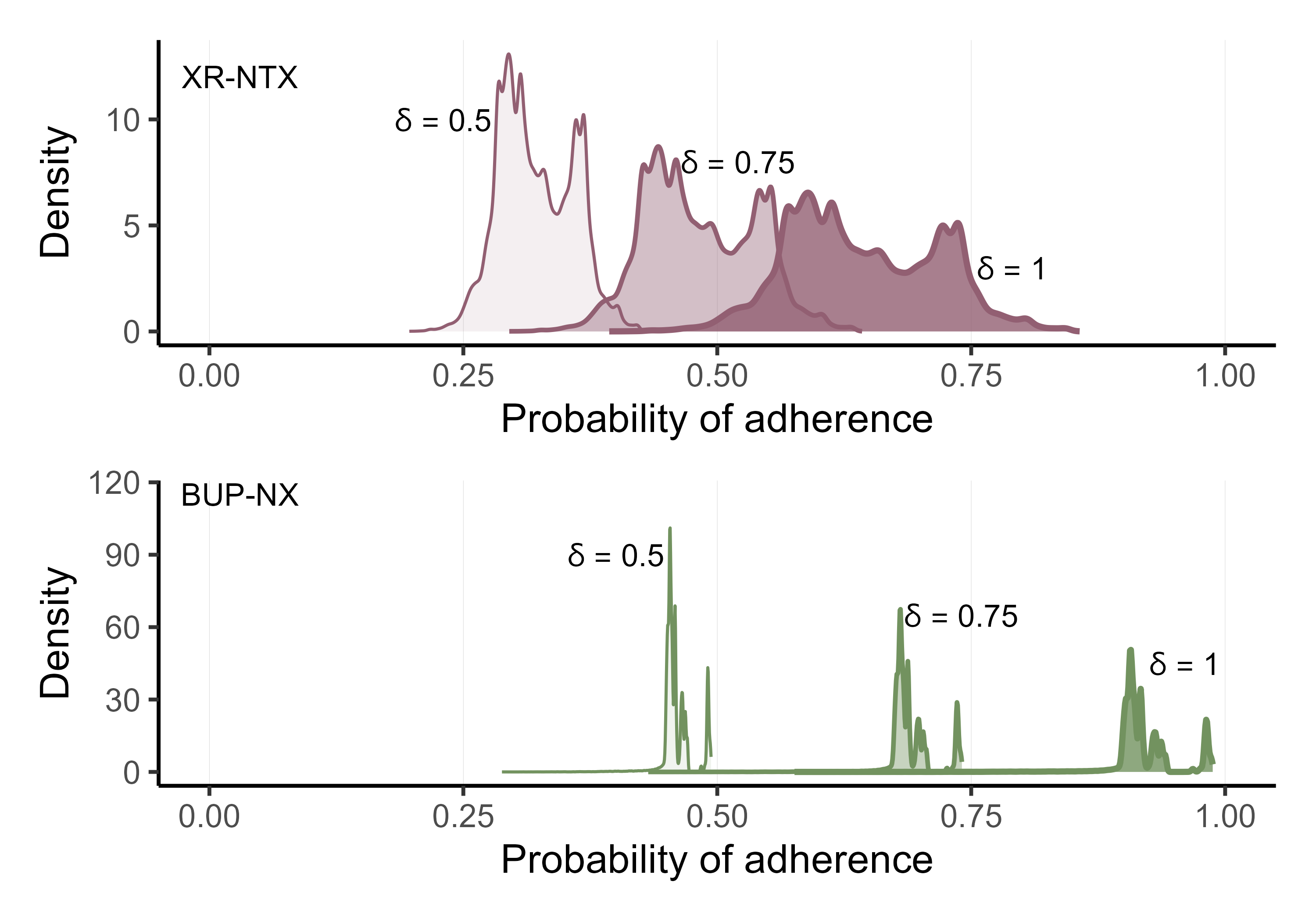}
\captionlistentry{}
\label{fig:disadh}
\end{figure}

\newpage
\noindent \textbf{Figure 3.} Scatterplots of 10,000 $\delta$ values for XR-NTX and BUP-NX drawn from each medication's trapezoidal probability distribution for $\delta$ (top) and the resulting predicted mean adherence probability to each medication for each draw in the TEDS-A cohort (bottom). Black circles mark draws where $\delta1 \leq \delta_0$ (i.e., comparing X:BOT to TEDS-A, the relative conditional adherence for BUP-NX is greater than that for XR-NTX). Gray circles mark draws where $\delta_1>\delta_0$.\\
\begin{figure}[h!]
\includegraphics[width=\textwidth]{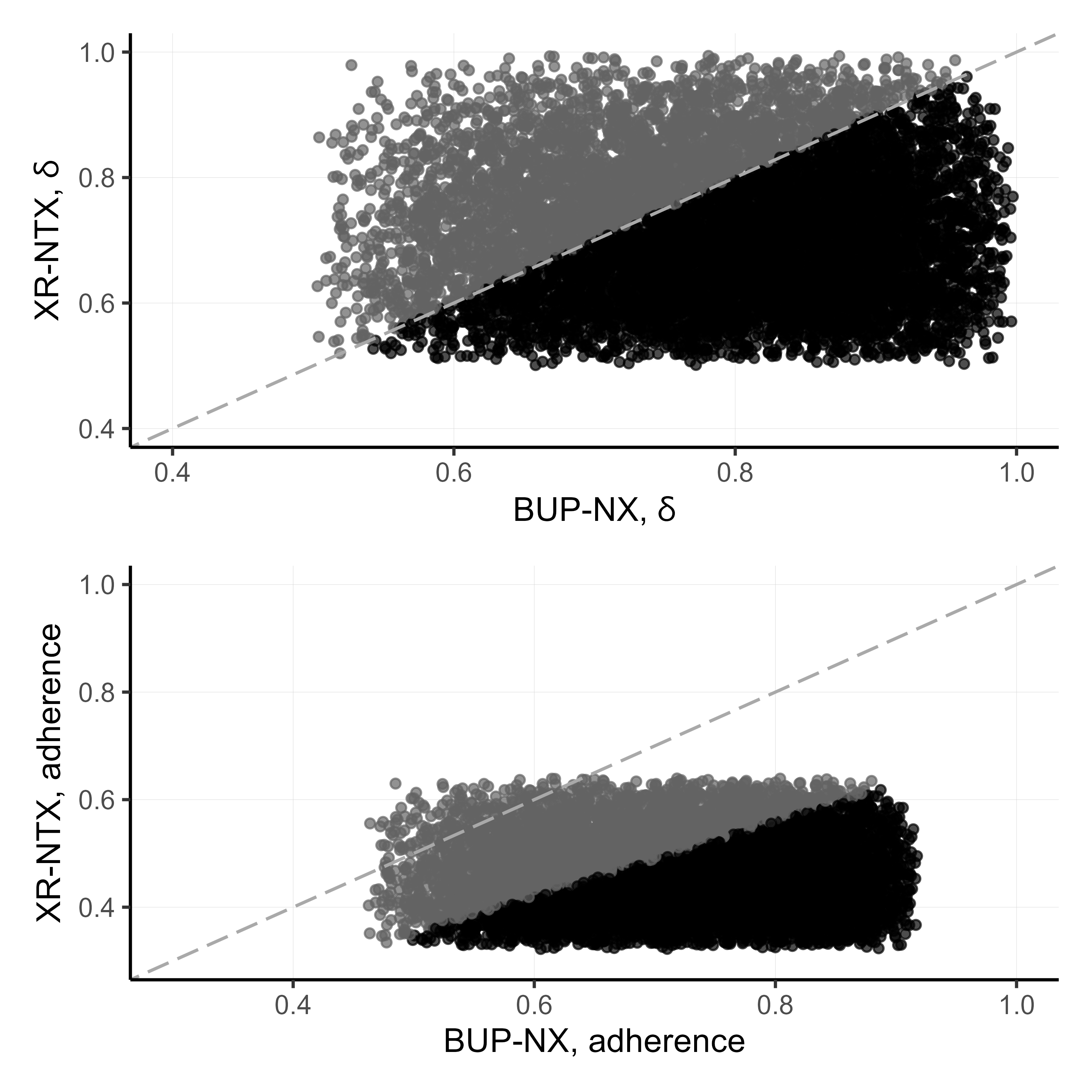}
\captionlistentry{}
\label{fig:draws}
\end{figure}

\newpage
\noindent \textbf{Figure 4.} Distribution of risk difference point estimates from 10000 draws of $\delta_0$ and $\delta_1$ from medication-specific trapezoidal probability distributions. Solid line marks distribution of all draws; dashed line marks distribution from subset of draws where $\delta1 \leq \delta_0$.\\
\begin{figure}[h!]
\includegraphics[width=\textwidth]{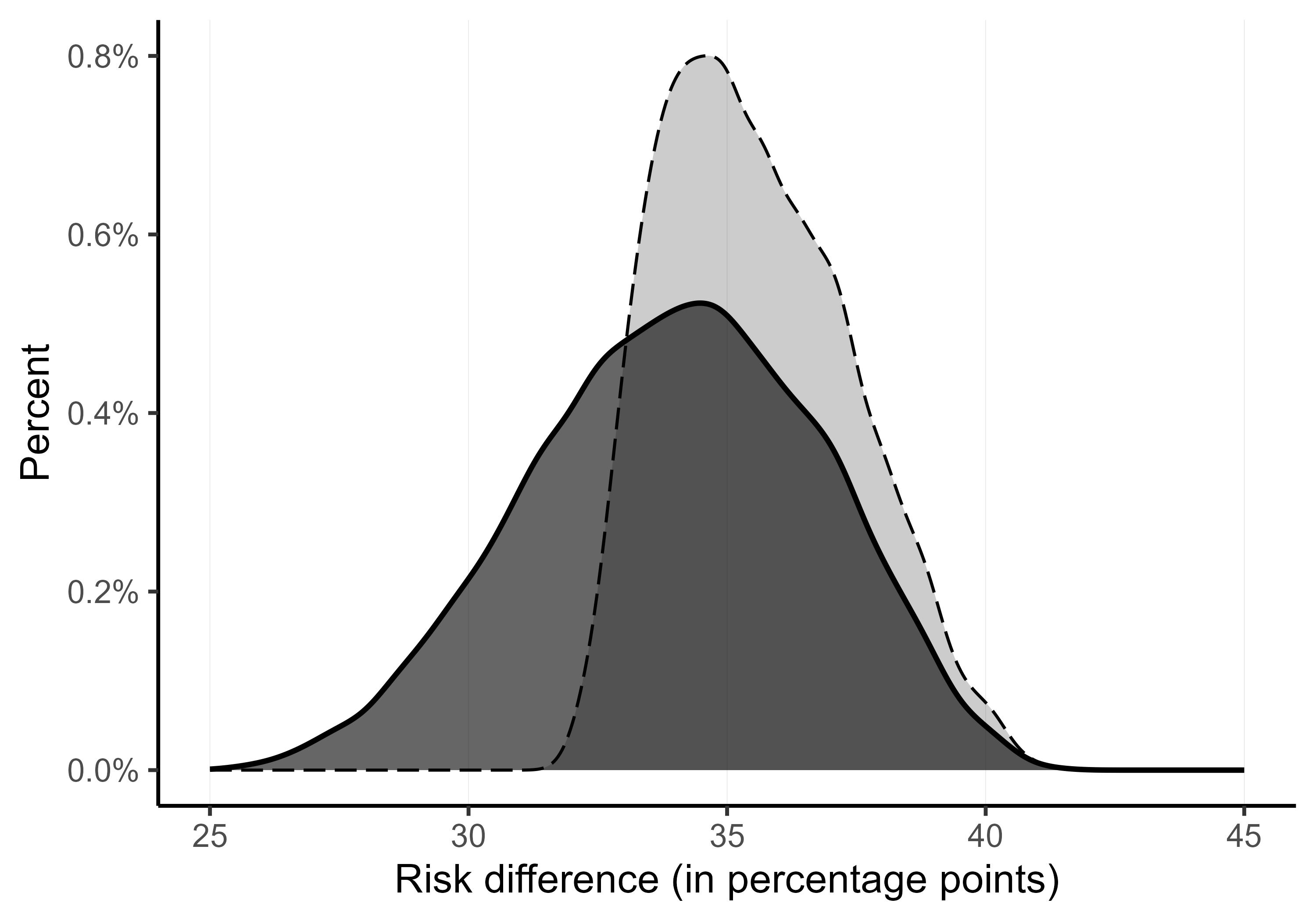}
\captionlistentry{}
\label{fig:drawsresults}
\end{figure}

\newpage

\setcounter{secnumdepth}{0}
\noindent \begin{center}{\LARGE Appendix for Transporting results from a trial to an external target population when trial participation impacts adherence}\end{center}%
\tableofcontents

\newpage
\section{Key proofs referenced in main text}

\subsection{Identification of $E[Y^a|S=0]$}
\subsubsection{Under setting (i) in Figure 1}
{\small
\begin{align*}
    E[Y^{a}&|S=0]\\
    &=E \left[ E\left(Y^{a}|W,S=0\right)|S=0\right] 
    \\
    &=E \left[ E\left(Y^{a}|W,S=1\right)|S=0\right] \text{  by A1 \& A2 }
    \\
    &=E \left[ E\left(Y^{a}|A=a,W,S=1\right)|S=0\right] \text{  by A3 \& A4} 
    \\
    &=E \left[ E\left(Y|A=a,W,S=1\right)|S=0\right] \text{  by A5}\\
\end{align*}}

\subsubsection{Under setting (ii) in Figure 1}
{\small
\begin{align*}
    E&[Y^{a}|S=0]\\
    &=E \bigg[ E\left(Y^{a}|W,S=0\right)|S=0 \bigg]
    \\
    &=E \bigg[ \sum_{z\in\{0,1\}} E\left(Y^{a}|W,Z^a=z,S=0\right) \Pr(Z^a=z|W,S=0) \bigg| S=0\bigg]
    \\
    &=E \bigg[ \sum_{z\in\{0,1\}} E\left(Y^{a}|W,Z^a=z,S=1\right) \Pr(Z^a=z|W,S=0) \bigg| S=0\bigg] \text{  by A1* \& A2*} 
    \\
    &=E \bigg[ \sum_{z\in\{0,1\}} E\left(Y^{a}|W,A=a,Z^a=z,S=1\right) \Pr(Z^a=z|W,S=0) \bigg| S=0\bigg] \text{  by A3* \& A4*}
    \\
    &\quad\quad\quad\quad\quad\quad\quad\quad \footnotesize{\text{A3* $\implies E\left(Y^{a}|W,Z^a=z,S=1\right)=E\left(Y^{a}|W,A=a,Z^a=z,S=1\right)$ (see next page)}}\\
    &=E \bigg[\big\{ E\left(Y^{a}|W,A=a,Z^a=1,S=1\right) \Pr(Z^a=1|W,S=1)\delta_a(W) + \\
    &\quad\quad \quad\quad E\left(Y^{a}|W,A=a,Z^a=0,S=1\right)(1-\Pr(Z^a=0|W,S=1)\delta_a(W)) \big\}\bigg| S=0\bigg] \text{  by definition of } \delta_a(W) 
    \\
    &=E \bigg[\big\{ E\left(Y^{a}|W,A=a,Z^a=1,S=1\right) \Pr(Z^a=1|W,A=a,S=1)\delta_a(W) + \\
    &\quad\quad \quad\quad E\left(Y^{a}|W,A=a,Z^a=0,S=1\right)(1-\Pr(Z^a=1|W,A=a,S=1)\delta_a(W))\big\}\bigg| S=0\bigg] \text{  by A3* \& A4*} 
    \\
    &\quad\quad\quad\quad\quad\quad\quad\quad \footnotesize{\text{A3* $\implies \Pr\left(Z^{a}=z|W,S=1\right)=\Pr\left(Z^{a}=z|W,A=a,S=1\right)$ (see next page)}}\\
    &=E \bigg[ \big\{E\left(Y|W,A=a,Z=1,S=1\right) \Pr(Z=1|W,A=a,S=1)\delta_a(W) + \\
    &\quad\quad \quad\quad E\left(Y|W,A=a,Z=0,S=1\right)(1-\Pr(Z=1|W,A=a,S=1)\delta_a(W))\big\}\bigg| S=0\bigg] \text{  by A5}
    \\
\end{align*}}%

\newpage
\subsubsection*{Assumptions implied by A3*: $(Y^a,Z^{a}) \indep A\mid W,S=1$}
\noindent 1. A3* $\implies \Pr(Y^a=z,Z^a=y|W,S=1)=\Pr(Y^a=y,Z^a=z|W,A=a,S=1)$\\
\noindent 2. A3* $\implies \Pr\left(Z^{a}=z|W,S=1\right)=\Pr\left(Z^{a}=z|W,A=a,S=1\right)$
\begin{align*}
    \sum_y \Pr(Y^a=y,Z^a=z|W,S=1) &= \sum_y \Pr(Y^a=y,Z^a=z|W,A=a,S=1)   \\
    \Pr(Z^a=z|W,S=1)  &= \Pr(Z^a=z|W,A=a,S=1)
\end{align*}

\noindent 3. A3* $\implies E(Y^a|W,Z^a,S=1)=E(Y^a|W,A=a,Z^a,S=1)$
\begin{align*}
    \Pr(Y^a=y|W,Z^a=z,S=1) &= \frac{\Pr(Y^a=y,Z^a=z|W,S=1)}{\Pr(Z^a=z|W,S=1)}  \\
    &= \frac{\Pr(Y^a=y,Z^a=z|W,A=a,S=1)}{\Pr(Z^a=z|W,A=a,S=1)}\text{ by A3*}\\
    &= \Pr(Y^a=y|W,A=a,Z^a=z,S=1)
\end{align*}

\noindent 4. A3* $\implies$ A3
\begin{align*}
    \sum_z \Pr(Y^a=y,Z^a=z|W,S=1) &= \sum_z \Pr(Y^a=y,Z^a=z|W,A=a,S=1)   \\
    \Pr(Y^a=y|W,S=1)  &= \Pr(Y^a=y|W,A=a,S=1)
\end{align*}

\subsubsection*{Assumption A4* implies A4}
\noindent A4*: $\Pr(A=a|W=w,Z^a=z,S=1)>0$ when $\Pr(S=0|W=w,Z^a=z)>0$\\
\noindent A4: $\Pr(A=a|W=w,S=1)>0$ when $\Pr(S=0|W=w)>0$\\
\noindent Marginalize each piece of A4* over $Z^a$\\

 $$\sum_z \Pr(A=a|W=w,Z^a=z,S=1)>0 = \Pr(A=a|W=w,S=1)>0$$
 $$\sum_z \Pr(S=0|W=w,Z^a=z)>0=\Pr(S=0|W=w)>0$$

\newpage
\subsection{Derivation of conjectured EIC}
For derivation we follow strategy discussed in reference [14] in which we use notation assuming $W$ is discrete. Let $\EIC(\cdot)$ to denote the operator that returns the EIC.
\begin{align*}
    \psi&(a,\delta_a)=\sum_w \left\{Q_{a,1}(w) m_a(w) \delta_a + Q_{a,0}(w)(1-m_a(w)\delta_a) \right\} \Pr(W=w|S=0) \\
\end{align*}

\noindent EIC of individual components
\begin{align*}
    &\EIC(Q_{a,1}(w))=\frac{\I(a,Z=1,w,S=1)}{\Pr(a,Z=1,w,S=1)}(Y-Q_{a,1}(w))\\
    &\EIC(Q_{a,0}(w))=\frac{\I(a,Z=0,w,S=1)}{\Pr(a,Z=0,w,S=1)}(Y-Q_{a,0}(w))\\
    &\EIC(m_a(w)\delta_a)=\delta_a\frac{\I(a,w,S=1)}{\Pr(a,w,S=1)}(\I(Z=1)-m_a(w))\\
    &\EIC(1-m_a(w)\delta_a)=-\delta_a\frac{\I(a,w,S=1)}{\Pr(a,w,S=1)}(\I(Z=1)-m_a(w))\\
    &\EIC(\Pr(W=w|S=0))=\frac{\I(S=0)}{\Pr(S=0)}(\I(W=w)-\Pr(W=w|S=0))
\end{align*}

\noindent Step 1
{\scriptsize
\begin{align*}
    \EIC&(Q_{a,1}(w) m_a(w) \delta_a + Q_{a,0}(w)(1-m_a(W)\delta_a)\\
    &=\EIC(Q_{a,1}(w))m_a(w) \delta_a + Q_{a,1}(w)\EIC(m_a(w)\delta_a) + \EIC(Q_{a,0}(w))(1-m_a(W)\delta_a) + Q_{a,0}(w)\EIC((1-m_a(W)\delta_a)\\
    &=\frac{\I(a,Z=1,w,S=1)}{\Pr(a,Z=1,w,S=1)}(Y-Q_{a,1}(w))m_a(w) \delta_a + Q_{a,1}(w)\delta_a\frac{\I(a,w,S=1)}{\Pr(a,w,S=1)}(\I(Z=1)-m_a(w)) \\
    &\quad\quad\quad\quad\quad + \frac{\I(a,Z=0,w,S=1)}{\Pr(a,Z=0,w,S=1)}(Y-Q_{a,0}(w))(1-m_a(W)\delta_a) - Q_{a,0}(w)\delta_a\frac{\I(a,w,S=1)}{\Pr(a,w,S=1)}(\I(Z=1)-m_a(w))\\
    &=\frac{\I(a,w,S=1)\I(Z=1)}{m_a(w)\Pr(a,w,S=1)}(Y-Q_{a,1}(w))m_a(w) \delta_a + Q_{a,1}(w)\delta_a\frac{\I(a,w,S=1)}{\Pr(a,w,S=1)}(\I(Z=1)-m_a(w)) \\
    &\quad\quad\quad\quad\quad + \frac{\I(a,w,S=1)\I(Z=0)}{(1-m_a(w))\Pr(a,w,S=1)}(Y-Q_{a,0}(w))(1-m_a(W)\delta_a) - Q_{a,0}(w)\delta_a\frac{\I(a,w,S=1)}{\Pr(a,w,S=1)}(\I(Z=1)-m_a(w))\\
    &=\frac{\I(a,w,S=1)}{\Pr(a,w,S=1)}\bigg\{\I(Z=1)(Y-Q_{a,1}(w)) \delta_a + \I(Z=1)Q_{a,1}(w)\delta_a-Q_{a,1}(w)m_a(w)\delta_a \\
    &\quad\quad\quad\quad\quad + \I(Z=0)(Y-Q_{a,0}(w))\frac{(1-m_a(w)\delta_a)}{(1-m_a(w))} - \I(Z=1)Q_{a,0}(w)\delta_a+Q_{a,0}(w)m_a(w)\delta_a\bigg\}\\
    &=\frac{\I(a,w,S=1)}{\Pr(a,w,S=1)}\bigg\{\I(Z=1)(Y-Q_{a,z}(w)) \delta_a + \I(Z=0)(Y-Q_{a,z}(w))\frac{(1-m_a(w)\delta_a)}{(1-m_a(w))}\\
    &\quad\quad\quad\quad\quad  + \I(Z=1)Q_{a,1}(w)\delta_a-Q_{a,1}(w)m_a(w)\delta_a - \I(Z=1)Q_{a,0}(w)\delta_a+Q_{a,0}(w)m_a(w)\delta_a\bigg\}\\
    &=\frac{\I(a,w,S=1)}{\Pr(a,w,S=1)}\bigg\{\bigg(\I(Z=1)\delta_a + \I(Z=0)\frac{(1-m_a(w)\delta_a)}{1-m_a(W)}\bigg)(Y-Q_{a,z}(w))  \\
    &\quad\quad\quad\quad\quad + \I(Z=1)Q_{a,1}(w)\delta_a-Q_{a,1}(w)m_a(w)\delta_a - \I(Z=1)Q_{a,0}(w)\delta_a+Q_{a,0}(w)m_a(w)\delta_a\bigg\}\\
    &=\frac{\I(a,w,S=1)}{\Pr(a,w,S=1)}\bigg\{\bigg(\I(Z=1)\delta_a + \I(Z=0)\frac{(1-m_a(w)\delta_a)}{1-m_a(W)}\bigg)(Y-Q_{a,z}(w))  + \delta_a(Q_{a,1}(w)- Q_{a,0}(w))(\I(Z=1)-m_a(w)) \bigg\}
   \end{align*}}

\noindent Step 2
{\scriptsize
\begin{align*}
    \EIC&(\left\{Q_{a,1}(w) m_a(w) \delta_a + Q_{a,0}(w)(1-m_a(w)\delta_a) \right\} \Pr(W=w|S=0))\\
    &=\EIC(Q_{a,1}(w) m_a(w) \delta_a + Q_{a,0}(w)(1-m_a(W)\delta_a) \Pr(W=w|S=0)) \\
    &\quad\quad\quad\quad\quad + \left\{Q_{a,1}(w) m_a(w) \delta_a + Q_{a,0}(w)(1-m_a(w)\delta_a) \right\}\EIC(\Pr(W=w|S=0))\\
    &=\frac{\I(a,w,S=1)\Pr(W=w|S=0)}{\Pr(a,w,S=1)}\bigg\{\bigg(\I(Z=1)\delta_a + \I(Z=0)\frac{(1-m_a(w)\delta_a)}{1-m_a(W)}\bigg)(Y-Q_{a,z}(w))  + \delta_a(Q_{a,1}(w)- Q_{a,0}(w))(\I(Z=1)-m_a(w)) \bigg\}\\
    &\quad\quad\quad\quad\quad + \left\{Q_{a,1}(w) m_a(w) \delta_a + Q_{a,0}(w)(1-m_a(w)\delta_a) \right\}\frac{\I(S=0)}{\Pr(S=0)}(\I(W=w)-\Pr(W=w|S=0))\\
    &=\frac{\I(a,w,S=1)(1-h(w))}{g_a(w)h(w)\Pr(S=0)}\bigg\{\bigg(\I(Z=1)\delta_a + \I(Z=0)\frac{(1-m_a(w)\delta_a)}{1-m_a(W)}\bigg)(Y-Q_{a,z}(w))  + \delta_a(Q_{a,1}(w)- Q_{a,0}(w))(\I(Z=1)-m_a(w)) \bigg\}\\
    &\quad\quad\quad\quad\quad + \frac{\I(S=0)}{\Pr(S=0)}\left\{Q_{a,1}(w) m_a(w) \delta_a + Q_{a,0}(w)(1-m_a(w)\delta_a) \right\}(\I(W=w)-\Pr(W=w|S=0))\\
    &=\frac{1}{\Pr(S=0)}\bigg[\frac{\I(a,w,S=1)(1-h(w))}{g_a(w)h(w)}\bigg\{\bigg(\I(Z=1)\delta_a + \I(Z=0)\frac{(1-m_a(w)\delta_a)}{1-m_a(W)}\bigg)(Y-Q_{a,z}(w))  + \delta_a(Q_{a,1}(w)- Q_{a,0}(w))(\I(Z=1)-m_a(w)) \bigg\}\\
    &\quad\quad\quad\quad\quad + \I(S=0)(\I(W=w)\left\{Q_{a,1}(w) m_a(w) \delta_a + Q_{a,0}(w)(1-m_a(w)\delta_a) \right\} \\
    &\quad\quad\quad\quad\quad -\left\{Q_{a,1}(w) m_a(w) \delta_a + Q_{a,0}(w)(1-m_a(w)\delta_a) \right\}\Pr(W=w|S=0)) \bigg]
\end{align*}}

\noindent Final step
{\scriptsize
\begin{align*}
    \EIC&\left(\sum_w\left\{Q_{a,1}(w) m_a(w) \delta_a + Q_{a,0}(w)(1-m_a(w)\delta_a) \right\} \Pr(W=w|S=0)\right)\\
    &=\sum_w\EIC(\left\{Q_{a,1}(w) m_a(w) \delta_a + Q_{a,0}(w)(1-m_a(w)\delta_a) \right\} \Pr(W=w|S=0))\\
    &=\frac{1}{\Pr(S=0)}\bigg[\sum_w\frac{\I(a,w,S=1)(1-h(w))}{g_a(w)h(w)}\\
    &\quad\quad\quad\quad\quad \bigg\{\bigg(\I(Z=1)\delta_a + \I(Z=0)\frac{(1-m_a(w)\delta_a)}{1-m_a(W)}\bigg)(Y-Q_{a,z}(w))  + \delta_a(Q_{a,1}(w)- Q_{a,0}(w))(\I(Z=1)-m_a(w)) \bigg\}\\
    &\quad\quad\quad\quad\quad + \I(S=0)\bigg\{\sum_w\I(W=w)(Q_{a,1}(w) m_a(w) \delta_a + Q_{a,0}(w)(1-m_a(w)\delta_a)) \bigg\} \\
    &\quad\quad\quad\quad\quad -\sum_w\left(Q_{a,1}(w) m_a(w) \delta_a + Q_{a,0}(w)(1-m_a(w)\delta_a) \right)\Pr(W=w|S=0)) \bigg]\\
    &=\frac{1}{\Pr(S=0)}\bigg[\frac{\I(a,S=1)(1-h(W))}{g_a(W)h(W)}\bigg\{\bigg(\I(Z=1)\delta_a + \I(Z=0)\frac{(1-m_a(W)\delta_a)}{1-m_a(W)}\bigg)\left(Y-Q_{a,z}(W)\right)  + \delta_a\left(Q_{a,1}(W)- Q_{a,0}(W)\right)\left(\I(Z=1)-m_a(W)\right) \bigg\}\\
    &\quad\quad\quad\quad\quad + \I(S=0)\bigg\{Q_{a,1}(W) m_a(W) \delta_a + Q_{a,0}(W)\left(1-m_a(W)\delta_a\right) -\psi(a,\delta_a) \bigg\} \bigg]
\end{align*}}

\newpage
\subsection{Asymptotic properties of one-step estimator}
\subsubsection{Rate double-robustness}

$$\sqrt{n}( \psi(\widehat P)-\psi(P)) \xrightarrow{d} N(0,\sigma^2)$$

\noindent By von Mises expansion
\begin{align*}
\psi (\widehat P) - \psi (P) &= -\underbrace{E\{\phi(O, \widehat P)\}}_{\text{``First order error''}} - \underbrace{R}_{\text{``Remainder''}}
\end{align*}

\noindent Conjectured EIC
{\scriptsize
$$\phi=\frac{S}{k}\frac{\I(a)}{g}\frac{1-h}{h}
\left(Z\delta + (1-Z)\frac{1-m\delta}{1-m} \right)
(Y - Q_{Z})  + \frac{S}{k}\frac{\I(a)}{g}\frac{1-h}{h}\delta(Q_{1} - Q_{0})(Z-m) 
+ \frac{1-S}{k} \mu- \frac{1-S}{k}\psi\\$$ 
}%
where $Q_1=E[Y|A=a,Z=1,W,S=1]$, $Q_0=E[Y|A=a,Z=0,W,S=1]$, $Q_Z=E[Y|A=a,Z,W,S=1]$, $m=\Pr(Z=1|A=a,W,S=1)$, $g=\Pr(A=a|W,S=1)$, $h=\Pr(S=1|W)$, $k=\Pr(S=0)$ and $\mu=Q_{1}m\delta - Q_{0}(1-m\delta)$\\~\\

\noindent Therefore,
{\scriptsize
\begin{align}
    R&=E[\phi(\widehat P)]+\psi(\widehat P)-\psi( P) \notag\\
    &=E\bigg[\frac{S}{\widehat k}\frac{\I(a)}{\widehat g}\frac{1-\widehat h}{\widehat h}
    \left(Z\delta + (1-Z)\frac{1-\widehat m\delta}{1-\widehat m} \right)(Y - \widehat Q_{Z}) +\frac{S}{\widehat k}\frac{\I(a)}{\widehat g}\frac{1-\widehat h}{\widehat h} \delta(\widehat Q_{1} - \widehat Q_{0})(Z-\widehat m) + \frac{1-S}{\widehat k}\widehat \mu- \frac{1-S}{\widehat k}\psi(\widehat P)   \bigg] + \psi(\widehat P)-\psi(P) \notag\\
    &=E\bigg[\frac{1-\widehat h}{\widehat k}\frac{g}{\widehat g}\frac{h}{\widehat h}
    \left(Z\delta + (1-Z)\frac{1-\widehat m\delta}{1-\widehat m} \right)(Q_Z - \widehat Q_{Z}) +\frac{1-\widehat h}{\widehat k}\frac{g}{\widehat g}\frac{h}{\widehat h} \delta(\widehat Q_{1} - \widehat Q_{0})(m-\widehat m)  + \frac{1-h}{\widehat k}\widehat \mu\bigg]- E\bigg[\frac{1-h}{\widehat k}\psi(\widehat P)   \bigg] + \psi(\widehat P)-\psi(P) \notag\\    
    &=E\bigg[\frac{gh}{\widehat g \widehat h}\frac{1-\widehat h}{\widehat k}
        \left(Z\delta + (1-Z)\frac{1-\widehat m\delta}{1-\widehat m} \right)(Q_Z - \widehat Q_{Z}) \textcolor{blue}{- \frac{1-\widehat h}{\widehat k}
        \left(Z\delta + (1-Z)\frac{1-\widehat m\delta}{1-\widehat m} \right)(Q_Z - \widehat Q_{Z}) + \frac{1-\widehat h}{\widehat k}
        \left(Z\delta + (1-Z)\frac{1-\widehat m\delta}{1-\widehat m} \right)(Q_Z - \widehat Q_{Z})}\notag\\
        &\quad\quad\quad +\frac{gh}{\widehat g \widehat h}\frac{1-\widehat h}{\widehat k} \delta(\widehat Q_{1} - \widehat Q_{0})(m-\widehat m) \textcolor{blue}{-\frac{1-\widehat h}{\widehat k} \delta(\widehat Q_{1} - \widehat Q_{0})(m-\widehat m)+\frac{1-\widehat h}{\widehat k} \delta(\widehat Q_{1} - \widehat Q_{0})(m-\widehat m)} \notag\\
        &\quad\quad\quad + \frac{1-h}{\widehat k}\widehat \mu \textcolor{blue}{-\frac{1-h}{\widehat k} \mu +\frac{1-h}{\widehat k} \mu} \bigg]  - \frac{k}{\widehat k}\psi(\widehat P)    + \psi(\widehat P)-\psi(P) \notag\\
    &=E\bigg[\textcolor{teal}{\left(\frac{gh}{\widehat g \widehat h}-1\right)\frac{1-\widehat h}{\widehat k}
        \left(Z\delta + (1-Z)\frac{1-\widehat m\delta}{1-\widehat m} \right)(Q_Z - \widehat Q_{Z})} + \frac{1-\widehat h}{\widehat k}
        \left(Z\delta + (1-Z)\frac{1-\widehat m\delta}{1-\widehat m} \right)(Q_Z - \widehat Q_{Z})\notag\\
        &\quad\quad\quad +\textcolor{brown}{\left(\frac{gh}{\widehat g \widehat h}-1 \right)\frac{1-\widehat h}{\widehat k} \delta(\widehat Q_{1} - \widehat Q_{0})(m-\widehat m)}  +\frac{1-\widehat h}{\widehat k} \delta(\widehat Q_{1} - \widehat Q_{0})(m-\widehat m) \notag\\
        &\quad\quad\quad + \frac{1-h}{\widehat k} (\widehat\mu -  \mu) \bigg] + \textcolor{violet}{\frac{k}{\widehat k} \psi(P)  - \frac{k}{\widehat k}\psi(\widehat P)    + \psi(\widehat P)-\psi( P) }\notag\\
    &=\textcolor{violet}{\frac{1}{\widehat k}\left(k-\widehat k\right)\left(\psi(P)-\psi(\widehat P) \right)} \label{tmp1}\\
        &\quad\quad\quad + E\bigg[\textcolor{teal}{\frac{1-\widehat h}{\widehat k \widehat g \widehat h}
        \left(Z\delta + (1-Z)\frac{1-\widehat m\delta}{1-\widehat m} \right)(gh-\widehat g \widehat h)(Q_Z - \widehat Q_{Z})}\bigg] \label{tmp2}\\
        &\quad\quad\quad + E\bigg[\textcolor{brown}{ \frac{1-\widehat h}{\widehat k \widehat g \widehat h} \delta(\widehat Q_{1} - \widehat Q_{0})(gh-\widehat g \widehat h)(m-\widehat m)}\bigg]\label{tmp3}\\
        &\quad\quad\quad + E\bigg[\frac{1-\widehat h}{\widehat k}
        \left(Z\delta + (1-Z)\frac{1-\widehat m\delta}{1-\widehat m} \right)(Q_Z - \widehat Q_{Z}) +\frac{1-\widehat h}{\widehat k} \delta(\widehat Q_{1} - \widehat Q_{0})(m-\widehat m)  + \frac{1-h}{\widehat k} (\widehat\mu -  \mu) \bigg]  \label{tmp4}
\end{align}}%

\noindent Focus on (\ref{tmp4})
{\scriptsize
\begin{align}
E\bigg[&\frac{1-\widehat h}{\widehat k}
        \left(Z\delta + (1-Z)\frac{1-\widehat m\delta}{1-\widehat m} \right)(Q_Z - \widehat Q_{Z})
        +\frac{1-\widehat h}{\widehat k} \delta(\widehat Q_{1} - \widehat Q_{0})(m-\widehat m) + \frac{1-h}{\widehat k}\left(\widehat Q_1 \widehat m \delta + \widehat Q_0 (1-\widehat m \delta) -  Q_1  m \delta -  Q_0 (1- m \delta) \right)\bigg] \notag\\
    &=E\bigg[\frac{1-\widehat h}{\widehat k}
        \left(Z\delta + (1-Z)\frac{1-\widehat m\delta}{1-\widehat m} \right)(Q_Z - \widehat Q_{Z})+\frac{1-\widehat h}{\widehat k} \delta(\widehat Q_{1} - \widehat Q_{0})(m-\widehat m) \notag\\
        &\quad\quad\quad+ \frac{1-h}{\widehat k}\left( \widehat m \delta (\widehat Q_1-\widehat Q_0)  \textcolor{blue}{- m \delta (\widehat Q_1-\widehat Q_0) + m \delta (\widehat Q_1-\widehat Q_0)}+ \widehat Q_0 -  Q_1  m \delta -  Q_0 + Q_0m \delta) \right)\bigg]\notag\\
    &=E\bigg[\frac{1-\widehat h}{\widehat k}
        \left(Z\delta + (1-Z)\frac{1-\widehat m\delta}{1-\widehat m} \right)(Q_Z - \widehat Q_{Z})+\frac{1-\widehat h}{\widehat k} \delta(\widehat Q_{1} - \widehat Q_{0})(m-\widehat m) \notag\\
        &\quad\quad\quad+ \frac{1-h}{\widehat k}\delta (\widehat Q_1-\widehat Q_0) (\widehat m - m) +  \frac{1-h}{\widehat k}\left((\widehat Q_1 -  Q_1  )m \delta + (\widehat Q_0-Q_0) (1-m \delta)  \right)\bigg]\notag\\
    &=E\bigg[\frac{1-\widehat h}{\widehat k}
        \left(Z\delta + (1-Z)\frac{1-\widehat m\delta}{1-\widehat m} \right)(Q_Z - \widehat Q_{Z})+\frac{h-\widehat h}{\widehat k} \delta(\widehat Q_{1} - \widehat Q_{0})(m-\widehat m)  +  \frac{1-h}{\widehat k}\left((\widehat Q_1 -  Q_1  )m \delta + (\widehat Q_0-Q_0) (1-m \delta)  \right)\bigg]\notag\\
    &=E\bigg[
    \frac{\delta(\widehat Q_{1} - \widehat Q_{0})}{\widehat k} (h-\widehat h)(m-\widehat m)\bigg] \label{tmp5}\\
        &\quad\quad\quad +E\bigg[
        \frac{1-\widehat h}{\widehat k}
        \left(Z\delta + (1-Z)\frac{1-\widehat m\delta}{1-\widehat m} \right)(Q_Z - \widehat Q_{Z}) 
         +  \frac{1-h}{\widehat k}\left((\widehat Q_1 -  Q_1  )m \delta + (\widehat Q_0-Q_0) (1-m \delta)  \right)\bigg] \label{tmp6}
\end{align}}%

\noindent Focus on (\ref{tmp6})

{\scriptsize
\begin{align}
E\bigg[&
        \frac{1-\widehat h}{\widehat k}
        \left(Z\delta + (1-Z)\frac{1-\widehat m\delta}{1-\widehat m} \right)(Q_Z - \widehat Q_{Z}) 
         +  \frac{1-h}{\widehat k}\left((\widehat Q_1 -  Q_1  )m \delta + (\widehat Q_0-Q_0) (1-m \delta)  \right)\bigg]\notag\\
&=E\bigg[
        \frac{1-\widehat h}{\widehat k}
        \left(Z\delta(Q_1 - \widehat Q_{1}) + (1-Z)\frac{1-\widehat m\delta}{1-\widehat m}(Q_0 - \widehat Q_{0})\right) 
         +  \frac{1-h}{\widehat k}\left((\widehat Q_1 -  Q_1  )m \delta + (\widehat Q_0-Q_0) (1-m \delta)  \right)\bigg]\notag\\
&=E\bigg[
        \frac{1-\widehat h}{\widehat k}m\delta(Q_1 - \widehat Q_{1}) + \frac{1-\widehat h}{\widehat k}(1-m)\frac{1-\widehat m\delta}{1-\widehat m}(Q_0 - \widehat Q_{0})  +  \frac{1-h}{\widehat k}(\widehat Q_1 -  Q_1  )m \delta + \frac{1-h}{\widehat k}(\widehat Q_0-Q_0) (1-m \delta)  \bigg]\notag\\
    &=E\bigg[
        \frac{1-\widehat h}{\widehat k}m\delta(Q_1 - \widehat Q_{1}) -  \frac{1-h}{\widehat k}( Q_1 -  \widehat Q_1  )m \delta\bigg] + E\bigg[  \frac{1-\widehat h}{\widehat k}(1-m)\frac{1-\widehat m\delta}{1-\widehat m}(Q_0 - \widehat Q_{0})   - \frac{1-h}{\widehat k}( Q_0-\widehat Q_0) (1-m \delta)  \bigg]\notag\\
    &=E\bigg[
        \frac{m\delta}{\widehat k}(h-\widehat h)(Q_1 - \widehat Q_{1})\bigg] \label{tmp7}\\
        &\quad\quad\quad +E\bigg[ \frac{1-\widehat h}{\widehat k}(1-\widehat m\delta)\frac{1-m}{1-\widehat m}(Q_0 - \widehat Q_{0})   - \frac{1-h}{\widehat k}( Q_0-\widehat Q_0) (1-m \delta)  \bigg] \label{tmp8}
\end{align}}%

\noindent Focus on (\ref{tmp8}) 

{\scriptsize
\begin{align}
E\bigg[& \frac{1-\widehat h}{\widehat k}(1-\widehat m\delta)\frac{1-m}{1-\widehat m}(Q_0 - \widehat Q_{0})   - \frac{1-h}{\widehat k}( Q_0-\widehat Q_0) (1-m \delta)  \bigg]\notag\\
    &=E\bigg[ \frac{1-\widehat h}{\widehat k}(1-\widehat m\delta)\frac{1-m}{1-\widehat m}(Q_0 - \widehat Q_{0}) \textcolor{blue}{-\frac{1-\widehat h}{\widehat k}(1-\widehat m\delta)(Q_0 - \widehat Q_{0})+\frac{1-\widehat h}{\widehat k}(1-\widehat m\delta)(Q_0 - \widehat Q_{0})}  - \frac{1-h}{\widehat k}( Q_0-\widehat Q_0) (1-m \delta)  \bigg]\notag\\
&=E\bigg[ \frac{1-\widehat m\delta}{1-\widehat m}\frac{1-\widehat h}{\widehat k}\left(\widehat m -m\right)(Q_0 - \widehat Q_{0}) \bigg] \label{tmp9}\\
&\quad\quad\quad+ E\bigg[\frac{1}{\widehat k}\left((1-\widehat h)(1-\widehat m\delta)  - (1-h)(1-m \delta)\right)( Q_0-\widehat Q_0)   \bigg]\label{tmp10}
\end{align}}%

$$R=(\ref{tmp1})+(\ref{tmp2})+(\ref{tmp3})+(\ref{tmp5})+(\ref{tmp7})+(\ref{tmp9})+(\ref{tmp10})$$

\noindent Each of the terms in the equation for the Remainder is a second-order term (i.e., it is a product of two errors)
\subsubsection{Double-robust consistency}
Each piece of the remainder term converges in probability to zero when the following models are correctly specified
\begin{align*}
    (\ref{tmp1}) &: \Pr(S=0)\\
    (\ref{tmp2}) &: (g_a(W),h(W))|Q_{a,z}(W)\\
    (\ref{tmp3}) &: (g_a(W),h(W))|m_a(W)\\
    (\ref{tmp5}) &: h(W)|m_a(W)\\
    (\ref{tmp7}) &: h(W)|Q_{a,z}(W)\\
    (\ref{tmp9}) &: m_a(W)|Q_{a,z}(W)\\
    (\ref{tmp10}) &: (h(W),m_a(W))|Q_{a,z}(W)\\
\end{align*}
\noindent Thus the remainder term will converge to zero under the following subsets of correctly specified models: 1) $Q_{a,z}(W)$ and $m_a(W)$; 2) $g_a(W)$, $h(W)$, and $m_a(W)$; or 3) $Q_{a,z}(W)$, $g_a(W)$, and $h(W)$.

\newpage
\section{Multiple adherence mediators or multi-time point adherence}
\begin{center}
DAG    
\begin{tikzpicture}
\tikzset{line width=1pt, outer sep=0pt, ell/.style={draw,fill=white, inner sep=2pt,line width=1pt}, swig vsplit={gap=5pt,inner line width right=0.5pt}};
\node[name=w, ell, shape=ellipse]{$W$};
\node[name=s, ell,above=8mm of w, shape=ellipse]{$S$};
\node[name=a, ell,right=5mm of w, shape=ellipse]{$A$};
\node[name=z1, ell,right=5mm of a, shape=ellipse]{$Z_1$};
\node[name=z2, ell,right=5mm of z1, shape=ellipse]{$Z_2$};
\node[name=y, ell,right=5mm of z2, shape=ellipse]{$Y$};
\draw[->,line width=1pt,>=stealth]
(w) edge (s)
(w) edge (a)
(s) edge (a)
(s) edge (z1)
(s) edge (z2)
(a) edge (z1)
(a) edge[out=-45,in=-145] (z2)
(a) edge[out=-45,in=-130] (y)
(z1) edge[out=-45,in=-145] (y)
(z2) edge (y)
(z1) edge (z2)
(w) edge[out=-60,in=-115] (y)
(w) edge[out=-45,in=-145] (z1)
(w) edge[out=-45,in=-130] (z2)
;
\end{tikzpicture}
SWIG
\begin{tikzpicture}
\tikzset{line width=1pt, outer sep=0pt, ell/.style={draw,fill=white, inner sep=2pt,line width=1pt}, swig vsplit={gap=5pt,inner line width right=0.5pt}};
\node[name=w, ell, shape=ellipse]{$W$};
\node[name=s, ell,above=8mm of w, shape=ellipse]{$S$};
\node[name=a, ell,right=5mm of w, shape=swig vsplit]{
    \nodepart{left}{$A$}
    \nodepart{right}{$a$}};
\node[name=z1, ell,right=5mm of a, shape=ellipse]{$Z^a_1$};
\node[name=z2, ell,right=5mm of z1, shape=ellipse]{$Z^a_2$};
\node[name=y, ell,right=5mm of z2, shape=ellipse]{$Y^a$};
\draw[->,line width=1pt,>=stealth]
(w) edge (s)
(w) edge (a)
(s) edge (a)
(s) edge (z1)
(s) edge (z2)
(a) edge (z1)
(a) edge[out=-45,in=-145] (z2)
(a) edge[out=-45,in=-130] (y)
(z1) edge[out=-45,in=-145] (y)
(z2) edge (y)
(z1) edge (z2)
(w) edge[out=-60,in=-115] (y)
(w) edge[out=-45,in=-145] (z1)
(w) edge[out=-45,in=-130] (z2)
;
\end{tikzpicture}
\end{center}
\noindent We define a $\delta$ parameter for each time point,
$$\Pr(Z_1^a=z_1|W,S=0)=\Pr(Z_1^a=z_1|W,S=1)\delta_{1,a,z_1}(W)$$
$$\Pr(Z_2^a=z_2|W,Z_1^a,S=0)=\Pr(Z_2^a=z_2|W,Z_1^a,S=1)\delta_{2,a,z_2}(W,Z_1)$$
\noindent Thus,
$$\Pr(Z_1^a=z_1,Z_2^a=z_2|W,S=0)=\Pr(Z_1^a=z_1,Z_2^a=z_2|W,S=1)\delta_{a,z_1,z_2}(W)$$
where $\delta_{a,z_1,z_2}(W)=\delta_{1,a,z_1}(W)\delta_{2,a,z_2}(W,Z_1)$\\
\noindent Assumptions
\begin{enumerate}
    \item[A1\textsuperscript{\textdagger}.] Mean conditional exchangeability over the samples:  $E(Y^a|W,Z_1^a,Z_2^a,S=1)-E(Y^a|W,Z_1^a,Z_2^a,S=0)$
    \item[A2\textsuperscript{\textdagger}.] Positivity of trial participation:  $\Pr(S=1|W=w,Z_1=z_1,Z_2=z_2)>0$ for all $a$ when $\Pr(S=0|W=w,Z_1=z_1,Z_2=z_2)>0$
    \item[A3\textsuperscript{\textdagger}.] Conditional exchangeability over treatment assignment in the trial:  $(Y^a,Z_1^a,Z_2^a) \indep A|W,S=1$. This assumption implies $\Pr(Z_1^a=z_1,Z_2^a=z_2|W,S=1)=\Pr(Z_1^a=z_1,Z_2^a=z_2|W,A=a,S=1)$ and $E(Y^a|W,Z_1^a,Z_2^a,L^a,S=1)=E(Y^a|W,A=a,Z_1^a,Z_2^a,L^a,S=1)$
    \item[A4\textsuperscript{\textdagger}.] Positivity of treatment assignment in the trial: $\Pr(A=a|W=w,Z_1^a=z_1,Z_2^a=z_2,S=1)>0$ when $\Pr(S=0|W=w,Z_1^a=z_1,Z_2^a=z_2)>0$
\end{enumerate}
{\scriptsize
\begin{align*}
    E&(Y^{a}|S=0)\\
    &=E \left[ E\left(Y^{a}|W,S=0\right)|S=0\right] \text{  by law of total probability}\\
    &=E \left[ \sum_{z_1,z_2} E\left(Y^{a}|W,Z_1^a=z_1,Z_2^a=z_2,S=0\right) \Pr(Z_1^a=z_1,Z_2^a=z_2|W,S=0) \bigg| S=0\right] \text{  by law of total probability}\\
    &=E \left[ \sum_{z_1,z_2} E\left(Y^{a}|W,Z_1^a=z_1,Z_2^a=z_2,S=1\right) \Pr(Z_1^a=z_1,Z_2^a=z_2|W,S=0) \bigg| S=0\right] \text{  by A1\textsuperscript{\textdagger} \& A2\textsuperscript{\textdagger}} 
    \\
    &=E \left[ \sum_{z_1,z_2} E\left(Y^{a}|W,A=a,Z_1^a=z_1,Z_2^a=z_2,S=1\right) \Pr(Z_1^a=z_1,Z_2^a=z_2|W,S=0) \bigg| S=0\right] \text{  by A3\textsuperscript{\textdagger} \& A4}
    \\
    &=E \left[ \sum_{z_1,z_2} E\left(Y^{a}|W,A=a,Z_1^a=z_1,Z_2^a=z_2,S=1\right) \Pr(Z_1^a=z_1,Z_2^a=z_2|W,S=1)\delta_{a,z_1,z_2}(W) \bigg| S=0\right] \text{  by definition of } \delta_{a,z_1,z_2}(W) 
    \\
    &=E \left[ \sum_{z_1,z_2} E\left(Y^{a}|W,A=a,Z_1^a=z_1,Z_2^a=z_2,S=1\right) \Pr(Z_1^a=z_1,Z_2^a=z_2|W,A=a,S=1)\delta_{a,z_1,z_2}(W) \bigg| S=0\right] \text{  by A3\textsuperscript{\textdagger} \& A4\textsuperscript{\textdagger}} 
    \\
    &=E \left[ \sum_{z_1,z_2} E\left(Y|W,A=a,Z_1=z_1,Z_2=z_2,S=1\right) \Pr(Z_1=z_1,Z_2=z_2|W,A=a,S=1)\delta_{a,z_1,z_2}(W) \bigg| S=0\right]  \text{  by A5}
\end{align*}}%

\newpage
\section{Incorporating post-treatment covariates that do not differ in distribution across $S$}
\noindent This setting is nearly identical to the prior page (where $L$ replaces $Z_1$ and $Z$ replaces $Z_2$)  except with one additional assumption: the distribution of $L$ does not differ across $S$ (articulated in new assumption A6 below), depicted on the diagram as the absence of an from $S$ to $L$.
\begin{center}
DAG    
\begin{tikzpicture}
\tikzset{line width=1pt, outer sep=0pt, ell/.style={draw,fill=white, inner sep=2pt,line width=1pt}, swig vsplit={gap=5pt,inner line width right=0.5pt}};
\node[name=w, ell, shape=ellipse]{$W$};
\node[name=s, ell,above=5mm of w, shape=ellipse]{$S$};
\node[name=a, ell,right=5mm of w, shape=ellipse]{$A$};
\node[name=z, ell,right=5mm of a, shape=ellipse]{$Z$};
\node[name=l, ell,above=5mm of z, shape=ellipse]{$L$};
\node[name=y, ell,right=5mm of z, shape=ellipse]{$Y$};
\draw[->,line width=1pt,>=stealth]
(w) edge (s)
(w) edge (a)
(s) edge (a)
(s) edge (z)
(a) edge (z)
(a) edge[out=-45,in=-145] (y)
(z) edge (y)
(w) edge[out=-60,in=-130] (y)
(w) edge[out=-45,in=-145] (z)
(a) edge (l)
(l) edge (z)
(l) edge (y)
;
\end{tikzpicture}
SWIG
\begin{tikzpicture}
\tikzset{line width=1pt, outer sep=0pt, ell/.style={draw,fill=white, inner sep=2pt,line width=1pt}, swig vsplit={gap=5pt,inner line width right=0.5pt}};
\node[name=w, ell, shape=ellipse]{$W$};
\node[name=s, ell,above=5mm of w, shape=ellipse]{$S$};
\node[name=a, ell,right=5mm of w, shape=swig vsplit]{
    \nodepart{left}{$A$}
    \nodepart{right}{$a$}};
\node[name=z, ell,right=5mm of a, shape=ellipse]{$Z^a$};
\node[name=l, ell,above=5mm of z, shape=ellipse]{$L^a$};
\node[name=y, ell,right=5mm of z, shape=ellipse]{$Y^{a}$};
\draw[->,line width=1pt,>=stealth]
(w) edge (s)
(s) edge (z)
(w) edge (a)
(s) edge (a)
(a) edge (z)
(a) edge[out=-45,in=-145] (y)
(z) edge (y)
(w) edge[out=-60,in=-130] (y)
(w) edge[out=-45,in=-160] (z)
(a) edge (l)
(l) edge (z)
(l) edge (y)
;
\end{tikzpicture}
\end{center}
We observe the data vector $(W_i,S_i=1,A_i,Z_i,L_i,Y_i)$ for individuals in the trial and $(W_i,S_i=0)$ for individuals in the target. We define a new parameter $\delta_a(W,L^a)$,
\begin{equation}
\Pr(Z^a=1|W,L^a,S=0)=\Pr(Z^a=1|W,L^a,S=1)\delta_a(W,L^a).
\end{equation} 
$\delta_a(W,L^a)$ can be interpreted as the ratio of the treatment-specific adherence in the target to adherence in the trial conditional on $W$ and $L^a$. Here we rely on $A1\textsuperscript{\textdagger}$-$A4\textsuperscript{\textdagger}$ (as on the previous page, rewritten below with $L$ in place of $Z_1$ and $Z$ in place of $Z_2$) and a new assumption A6.
\begin{enumerate}
    \item[A1\textquotesingle.] Mean conditional exchangeability over the samples: $E(Y^a|W,Z^a,L^a,S=1)-E(Y^a|W,Z^a,L^a,S=0) $
    \item[A2\textquotesingle.] Positivity of trial participation: $\Pr(S=1|W=w,L^a=l,Z^a=z)>0$ for all $a$ when $\Pr(S=0|W=w,L^a=l,Z^a=z)>0$
    \item[A3\textquotesingle.] Conditional exchangeability over treatment assignment in the trial: $(Y^a,L^a,Z^a) \indep A|W,S=1$. This assumption implies $E(Z^a|W,L^a,S=1)=E(Z^a|W,A=a,L^a,S=1)$ and $E(Y^a|W,Z^a,L^a,S=1)=E(Y^a|W,A=a,Z^a,L^a,S=1)$  
    \item[A4\textquotesingle.] Positivity of treatment assignment in the trial: $\Pr(A=a|W=w,L^a=l,Z^a=z,S=1)>0$ when $\Pr(S=0|W=w,L^a=l,Z^a=z)>0$
    \item[A6.] Conditional exchangeability over the samples: participation in the trial is independent of the potential covariate $L^a$, conditional on baseline covariates, $S\indep L^a|W$. 
\end{enumerate}
{\scriptsize
\begin{align*}
    E&(Y^{a}|S=0)\\
    &=E \left[ E\left[E\left(Y^{a}|W,L^a,S=0\right)|W,S=0\right]|S=0\right] \text{  by law of total probability}\\
    &=E \left[ E\left[ \sum_{z\in\{0,1\}}E\left(Y^{a}|W,Z^a=z,L^a,S=0\right)\Pr(Z^a=z|W,L^a,S=0)\bigg| W,S=0\right]\bigg| S=0\right] \text{  by law of total probability}\\
    &=E \left[ E\left[ \sum_{z\in\{0,1\}}E\left(Y^{a}|W,Z^a=z,L^a,S=1\right)\Pr(Z^a=z|W,L^a,S=0)\bigg| W,S=0\right]\bigg| S=0\right]  \text{  by A1\textquotesingle\space \& A2\textquotesingle\space} 
    \\
    &=E \left[ E\left[ \sum_{z\in\{0,1\}}E\left(Y^{a}|W,A=a,Z^a=z,L^a,S=1\right)\Pr(Z^a=z|W,L^a,S=0)\bigg| W,S=0\right]\bigg| S=0\right]  \text{  by A3\textquotesingle\space \& A4\textquotesingle\space}
    \\
    &=E \big[E\big[E\left(Y^{a}|W,A=a,Z^a=1,L^a,S=1\right) \Pr(Z^a=1|W,L^a,S=1)\delta_a(W,L^a) + \\
    &\quad\quad E\left(Y^{a}|W,A=a,Z^a=0,L^a,S=1\right)(1-\Pr(Z^a=1|W,L^a,S=1)\delta_a(W,L^a))  \big| W,S=0\big]\big| S=0\big] \text{  by definition of } \delta_a(W,L^a) 
    \\
    &=E \big[E\big[E\left(Y^{a}|W,A=a,Z^a=1,L^a,S=1\right) \Pr(Z^a=1|W,L^a,S=1)\delta_a(W,L^a) + \\
    &\quad\quad E\left(Y^{a}|W,A=a,Z^a=0,L^a,S=1\right)(1-\Pr(Z^a=1|W,L^a,S=1)\delta_a(W,L^a))  \big| W,S=1\big]\big| S=0\big] \text{  by A6 \& A2\textquotesingle} 
    \\
    &=E \big[E\big[E\left(Y^{a}|W,A=a,Z^a=1,L^a,S=1\right) \Pr(Z^a=1|W,A=a,L^a,S=1)\delta_a(W,L^a) + \\
    &\quad\quad E\left(Y^{a}|W,A=a,Z^a=0,L^a,S=1\right)(1-\Pr(Z^a=1|W,A=a,L^a,S=1)\delta_a(W,L^a))  \big| W,A=a,S=1\big]\big| S=0\big]  \text{  by A3\textquotesingle\space  \& A4\textquotesingle\space}
    \\
    &=E \big[E\big[E\left(Y|W,A=a,Z=1,L,S=1\right) \Pr(Z=1|W,A=a,L,S=1)\delta_a(W,L^a) + \\
    &\quad\quad E\left(Y^{a}|W,A=a,Z=0,L^a,S=1\right)(1-\Pr(Z=1|W,A=a,L,S=1)\delta_a(W,L^a))  \big| W,A=a,S=1\big]\big| S=0\big]  \text{  by A5}
\end{align*}}%

\newpage
\section{Estimating equations for sandwich variance estimator}
\subsection{G-computation estimator}
$$\widehat\psi_{G}(a,\delta_a)=
\frac{1}{n_0}
\sum_{i=1}^n \I(S_i=0) \left\{\widehat Q_{a,1}(W_i) \widehat m_a(W_i) \delta_a + \widehat Q_{a,0}(W_i)(1- \widehat m_a(W_i) \delta_a)\right\}$$

\noindent Consider a binary outcome and a binary adherence indicator modeled using logistic regression. All models are restricted to individuals with $A=a$ (i.e., stratified on treatment assignment). The outcome models are fit separately by adherence. Let $X_i=(1,W_i)$, the $i$th row of the design matrix; $\beta_1$ and $\beta_0$ be the (column) vectors of the parameters of the outcome models where $Z=1$ and $Z=0$, respectively; and $\alpha$ be the (column) vector of parameters of the adherence model. The estimating equations are,
$$\begin{bmatrix}
\I(A_i=a)Z_iS_i(Y_i - \expit(X_i \beta_1))X_i \\
\I(A_i=a)(1-Z_i)S_i(Y_i - \expit(X_i \beta_0))X_i \\
\I(A_i=a)S_i(Z_i - \expit(X_i \alpha))X_i \\
(1-S_i)(\widehat Y^{*}_i - \psi_{G}(a,\delta_a))
\end{bmatrix}$$
where $\widehat Y^{*}_i = \expit(X_i \widehat \beta_1)\expit(X_i \widehat \alpha) \delta_a + \expit(X_i \widehat \beta_0) (1-\expit(X_i \widehat \alpha) \delta_a) $  

\subsection{One-step estimator}
{\small
\begin{align*}
\widehat\psi_{OS}&(a,\delta_a)\\
&=\frac{1}{n_0}\sum_{i=1}^n 
\Bigg[\frac{\I(A_i=a,S_i=1)}{\widehat g_a(W_i)}\frac{1-\widehat h(W_i)}{\widehat h(W_i)}\\
&\quad\quad\quad\quad\quad\Bigg\{
\left(\I(Z_i=1)\delta_a + \I(Z_i=0)\frac{1-\widehat 
 m_a(W_i)\delta_a}{1-\widehat  m_a(W_i)} \right)
(Y - \widehat  Q_{a,Z_i}(W_i))- \delta_a\left(\widehat  Q_{a,1}(W_i) - \widehat Q_{a,0}(W_i)\right)\left(Z_i-\widehat  m_a(W_i)\right) \Bigg\}\\
&\quad\quad\quad\quad\quad\quad\quad+ \I(S_i=0)\left\{ \widehat Q_{a,1}(W_i)\widehat m_a(W_i)\delta_a + \widehat Q_{a,0}(W_i)(1- \widehat m_a(W_i)\delta_a)\right\}\Bigg]
\end{align*}}%

\noindent Consider a binary outcome and a binary adherence indicator modeled using logistic regression. The outcome and adherence models are restricted to individuals with $A=a$ (i.e., stratified on treatment assignment). The outcome models are fit separately by adherence. Let $X_i=(1,W_i)$, the $i$th row of the design matrix; $\beta_1$ and $\beta_0$ be the (column) vectors of the parameters of the outcome models where $Z=1$ and $Z=0$, respectively; $\alpha$ be the (column) vector of parameters of the adherence model; $\gamma$ be the (column) vector of parameters of the treatment model; and $\epsilon$ be the (column) vector of parameters of the selection model. The estimating equations are,
$$\begin{bmatrix}
\I(A_i=a)Z_iS_i(Y_i - \expit(X_i \beta_1))X_i \\
\I(A_i=a)(1-Z_i)S_i(Y_i - \expit(X_i \beta_0))X_i \\
\I(A_i=a)S_i(Z_i - \expit(X_i \alpha))X_i \\
S_i(A_i - \expit(X_i \gamma))X_i \\
(S_i - \expit(X_i \epsilon))X_i \\
\widehat Y^{\dag}_i - (1-S_i)\psi_{OS}(a,\delta_a)
\end{bmatrix}$$
where {\footnotesize
\begin{align*}
\widehat Y^\dag_i&=\frac{\I(A_i=a,S_i=1)}{\expit(X_i \widehat \gamma)}\frac{1-\expit(X_i \widehat \epsilon)}{\expit(X_i \widehat \epsilon)}\\
&\quad\quad\quad\Bigg\{
\left(\I(Z_i=1)\delta_a + \I(Z_i=0)\frac{1-
\expit(X_i \widehat \alpha)\delta_a}{1-\expit(X_i \widehat \alpha)} \right)
(Y - Z_i\expit(X_i \widehat \beta_1)-(1-Z_i)\expit(X_i \widehat \beta_0)) \\
&\quad\quad\quad\quad\quad\quad- \delta_a\left(\expit(X_i \widehat \beta_1) - \expit(X_i \widehat \beta_0)\right) \left(Z_i - \expit(X_i \widehat \alpha)\right)\Bigg\}\\
&\quad\quad\quad\quad+ \I(S_i=0)\left\{ \expit(X_i \widehat \beta_1)\expit(X_i \widehat \alpha)\delta_a + \expit(X_i \widehat \beta_0)(1- \expit(X_i \widehat \alpha)\delta_a)\right\}
\end{align*}}%

\newpage
\section{Previously published one-step estimators used in the application}
\subsection{For $E[Y^a|S=1]$}
Let $\widehat \theta(a)_{OS}$ be the one-step estimator of $E[Y^a|S=1]=E[E[Y|A=s,W,S=1]|S=1]$ and $Q^{'}_a(W_i)$ be the $W$-conditional outcome expectation when $A=a$ in the trial (note that this is not conditional on $Z$, in contrast to $Q$ in the main text). 
{\small
\begin{align*}
\widehat\theta_{OS}(a)=\frac{1}{n_1}\sum_{i=1}^n 
\Bigg[\frac{\I(A_i=a,S_i=1)}{\widehat g_a(W_i)}\left(Y - \widehat  Q^{'}_{a}(W_i)\right) + \I(S_i=1) \widehat Q^{'}_{a}(W_i)\Bigg].
\end{align*}}%

\subsection{For $E[Y^a|S=0]$ under setting (i) in Figure 1}
Let $\widehat \theta(a)^{'}_{OS}$ be the one-step estimator of $E[Y^a|S=0]=E[E[Y|A=s,W,S=1]|S=0]$. 
{\small
\begin{align*}
\widehat\theta^{'}_{OS}(a)=\frac{1}{n_0}\sum_{i=1}^n 
\Bigg[\frac{\I(A_i=a,S_i=1)}{\widehat g_a(W_i)}\frac{1-\widehat h(W_i)}{\widehat h(W_i)}
\left(Y - \widehat  Q^{'}_{a}(W_i)\right) + \I(S_i=0) \widehat Q^{'}_{a}(W_i)\Bigg].
\end{align*}}%

\newpage
\section{Appendix Figures}
\subsection{Appendix Figure 1. \text{Trapezoidal probability distributions for $\delta_0$ and $\delta_1$}}
\includegraphics[width=\textwidth]{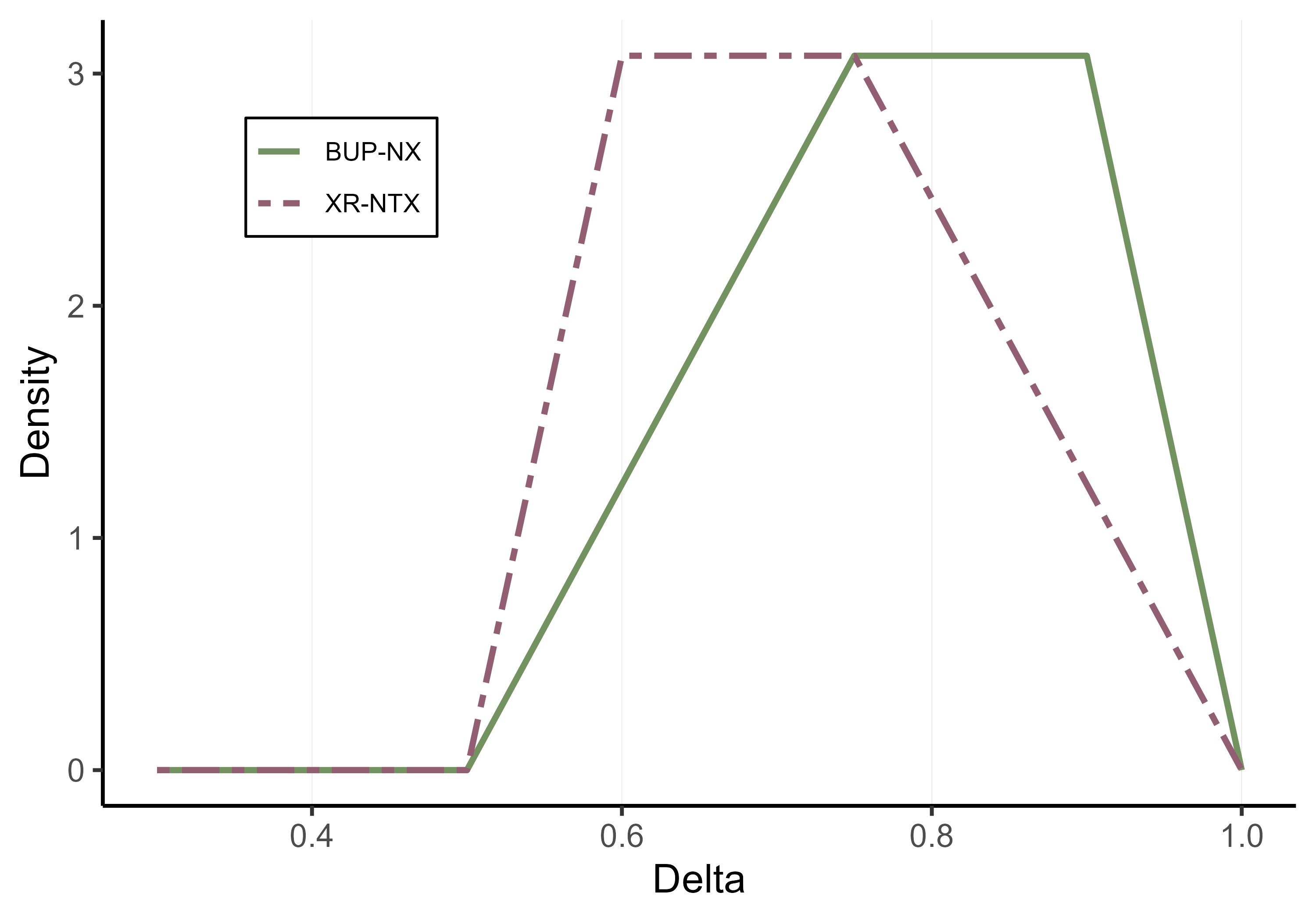}


\newpage
\section{Appendix Tables}
\subsection{Appendix Table 1. \text{Missingness in TEDS-A cohort}}
\begin{longtable}{lc}
\toprule
\textbf{Characteristic} & \textbf{Missingness}\\
&N = 446,210 \\ 
\midrule
Age & 0 (0\%) \\ 
Male & 0 (0\%) \\ 
Race & 6,746 (1.5\%) \\ 
Hispanic/Latine & 6,665 (1.5\%) \\ 
Education & 15,734 (3.5\%) \\ 
Unemployed & 11,062 (2.5\%) \\ 
Homeless & 14,102 (3.2\%) \\ 
Intravenous use & 0 (0\%) \\ 
Age at first use & 2,942 (0.7\%) \\ 
Cannabis & 0 (0\%) \\ 
Cocaine/crack & 0 (0\%) \\ 
Amphetamines & 0 (0\%) \\ 
Sedatives & 0 (0\%) \\ 
Any missing & 36,917 (8.3\%) \\ 
\bottomrule
\end{longtable}

\newpage
\subsection{Appendix Table 2. \text{Geographies included in TEDS-A cohort in each year}}
{\small
\begin{longtable}{lrrr}
\toprule
State & 2014 & 2015 & 2016 \\ 
\midrule
Alabama & 1 & 1 & 1 \\ 
Alaska & 1 & 1 & 1 \\ 
Arizona & 1 & 1 & 1 \\ 
Arkansas & 1 & 1 & 1 \\ 
California & 1 & 1 & 1 \\ 
Colorado & 1 & 1 & 1 \\ 
Connecticut & 1 & 1 & 1 \\ 
DC & 1 & 1 & 1 \\ 
Delaware & 1 & 1 & 1 \\ 
Florida & 1 & 1 & 1 \\ 
Georgia & 1 & 1 & 1 \\ 
Hawaii & 1 & 1 & 1 \\ 
Idaho & 1 & 1 & 1 \\ 
Illinois & 1 & 1 & 1 \\ 
Indiana & 1 & 1 & 1 \\ 
Iowa & 1 & 1 & 1 \\ 
Kansas & 1 & 1 & 1 \\ 
Kentucky & 1 & 1 & 1 \\ 
Louisiana & 1 & 1 & 1 \\ 
Maine & 1 & 1 & 1 \\ 
Maryland & 1 & 1 & 1 \\ 
Massachusetts & 1 & 1 & 1 \\ 
Michigan & 1 & 1 & 1 \\ 
Minnesota & 0 & 0 & 0 \\ 
Mississipi & 1 & 1 & 1 \\ 
Missouri & 1 & 1 & 1 \\ 
Montana & 1 & 1 & 1 \\ 
Nebraska & 1 & 1 & 1 \\ 
Nevada & 1 & 1 & 1 \\ 
New Hampshire & 1 & 0 & 0 \\ 
New Jersey & 1 & 1 & 1 \\ 
New Mexico & 1 & 1 & 0 \\ 
New York & 1 & 1 & 1 \\ 
North Carolina & 1 & 1 & 1 \\ 
North Dakota & 1 & 1 & 1 \\ 
Ohio & 1 & 1 & 1 \\ 
Oklahoma & 1 & 1 & 1 \\ 
Oregon & 1 & 0 & 0 \\ 
Pennsylvania & 1 & 1 & 1 \\ 
Puerto Rico & 1 & 0 & 1 \\ 
Rhode Island & 1 & 1 & 1 \\ 
South Carolina & 0 & 0 & 1 \\ 
South Dakota & 0 & 0 & 0 \\ 
Tennessee & 1 & 1 & 1 \\ 
Texas & 1 & 1 & 1 \\ 
Utah & 1 & 1 & 1 \\ 
Vermont & 1 & 1 & 1 \\ 
Virginia & 1 & 1 & 1 \\ 
Washington & 0 & 0 & 1 \\ 
West Virginia & 1 & 1 & 0 \\ 
Wisconsin & 1 & 1 & 1 \\ 
Wyoming & 1 & 1 & 1 \\ 
Total & 48 & 45 & 46 \\ 
\bottomrule
\end{longtable}}

\newpage
\subsection{Appendix Table 3. \text{Coding of covariates}}
\begin{longtable}{l L{12cm}}
\multicolumn{2}{l}{Sociodemographics} \\ 
\midrule
Age & \multicolumn{1}{c}{Continuous}\\
& TEDS-A: Raw files provide categorical age (AGE). Continuous age values were imputed as the mean of the age category  \\ 
Male & \multicolumn{1}{c}{Categorical: Yes, No}  \\ 
& TEDS-A: Recoded from GENDER \\
Race &  \multicolumn{1}{c}{Categorical: Black, White, Other}  \\ 
& TEDS-A: Coded from RACE: 4=Black; 5=White; {1,2,3,9,6,8,7}=Other  \\
Hispanic/Latine & \multicolumn{1}{c}{Categorical: Yes, No}  \\
& TEDS-A: Coded from ETHNIC: {1,2,3,5}=Yes; 4=No  \\
Education & \multicolumn{1}{c}{Categorical: <HS, HS/GED, HS+}  \\
& TEDS-A: Coded from EDUC: {1,2}=<HS; {3}=HS/GED; {4,5}=HS+  \\
Unemployed & \multicolumn{1}{c}{Categorical: Yes, No}  \\
& TEDS-A: Coded from EMPLOY: {3,4}=Yes; {1,2}=No  \\
Homeless & \multicolumn{1}{c}{Categorical: Yes, No}  \\
& TEDS-A: Coded from LIVARAG: {1}=Yes; {2,3}=No\\
\midrule
\multicolumn{2}{l}{Opioid use} \\ 
\midrule
Intravenous use & \multicolumn{1}{c}{Categorical: Yes, No}  \\
& TEDS-A: Coded from IDU: 1=Yes; 0=No \\
Age at first use &  \multicolumn{1}{c}{Categorical: <15, 15-20, 20-29, 30+}  \\
& TEDS-A: Coded from FIRSTUSE variables when SUB marked as opioid (SUB={5,6,7}): {1,2}=<15; {3,4}=15-20; {5,6}:21-29; 7=30+   \\
\midrule
\multicolumn{2}{l}{Other substance use} \\ 
\midrule
Cannabis & \multicolumn{1}{c}{Categorical: Yes, No}  \\
& TEDS-A: Coded from MARFLG indicating that marijuana recorded as primary, secondary, or tertiary substance  \\
Cocaine/crack & \multicolumn{1}{c}{Categorical: Yes, No}  \\
& TEDS-A: Coded from COKEFLG indicating that cocaine or crack recorded as primary, secondary, or tertiary substance \\
Amphetamines & \multicolumn{1}{c}{Categorical: Yes, No}  \\
& TEDS-A: Coded from MTHAMFLG, AMPHFLG, STIMFLG indicating that amphetamine, methamphetamine, or other stimulant recorded as primary, secondary, or tertiary substance  \\
Sedatives & \multicolumn{1}{c}{Categorical: Yes, No}  \\
& TEDS-A: Coded from BENZFLG, TRNQFLG, BARBFLG, SEDHFLG indicating that benzodiazepine, other tranquilizer, barbiturate, or other sedative/hypnotic recorded as primary, secondary, or tertiary substance \\
\bottomrule
\end{longtable}


\end{document}